\documentstyle[preprint,eqsecnum,aps]{revtex}

\draft

\date{\today}

\begin{document}

\draft

\title{Extended Classical Over--Barrier Model for Collisions of Highly
  Charged Ions with Conducting and Insulating Surfaces}

\author{Jens J. Ducr\'ee\footnote{permanent address: Institut f\"ur
    Kernphysik, Wilhelm Klemm--Str. 9, D--48149 M\"unster}, Fulvio
  Casali, and Uwe Thumm\thanks{corresponding author}}

\address{J. R. Macdonald Laboratory, Department of Physics, Kansas
  State University, Manhattan, Kansas 66506-2604, USA}

\date{\today }

\maketitle

\begin{abstract}
  We have extended the classical over--barrier model to simulate the
  neutralization dynamics of highly charged ions interacting under
  grazing incidence with conducting and insulating surfaces. Our
  calculations are based on simple model rates for resonant and Auger
  transitions. We include effects caused by the dielectric response of
  the target and, for insulators, localized surface charges.
  Characteristic deviations regarding the charge transfer processes
  from conducting and insulating targets to the ion are discussed. We
  find good agreement with previously published experimental data for
  the image energy gain of a variety of highly charged ions impinging
  on Au, Al, LiF and KI crystals.

\end{abstract}

\pacs{PACS numbers: 34.50.Dy, 34.70.+e, 79.20.Rf}

\section{Introduction}
\label{sec:intro}

Within the past decade, a rapidly increasing number of research
projects has been devoted to the investigation of interactions between
highly charged ions (HCI) and surfaces (for recent reviews see
\cite{And91,Bur93,Aum95}). These activities are of importance for
present and future applications, such as semiconductor fabrication,
nanostructure technology, and surface chemistry. They are also of
interest to basic research due to the challenging interplay of
fundamental electronic interactions to be considered in the detailed
understanding of the highly complex interaction dynamics. By now, a
certain level of consent has emerged with regard to charge exchange
and ionization processes that take place before an HCI gets in close
contact with a metal target surface, and a mainly classical approach,
the "classical over-barrier model" (COM), first presented by
Burgd\"orfer, Lerner, and Meyer \cite{Bur91,Bur95}, was found to
adequately represent the most important physical aspects of the
electron capture, recapture, and emission sequence
\cite{Bur95,Bar95,Thu95A,Thu95B,Bur96,Lem96,Thu97}.

Typically, an incident HCI captures several conduction band electrons
at large distances from the surface into highly excited states which
leads to the temporary formation of a "hollow ion" in front of the
surface. At ion--surface distances that are smaller than or about
equal to the classical radii of active HCI orbitals, the theoretical
description becomes more difficult due to the strong perturbation of
the initial electron distribution of the surface and the intricate
molecular dynamics involved. For this reason, most first--principle
calculations have been applied to incident ions in low charge states
\cite{Bur87,Thu89,Bor93,Deu97}. For higher incident charge states the
detailed quantum mechanical treatment is complicated by a large number
of ionic states that are energetically degenerate with the target
conduction band, and a first--principle approach remains a formidable
task \cite{Wil95,Bor96,Kur96,Nor96}.

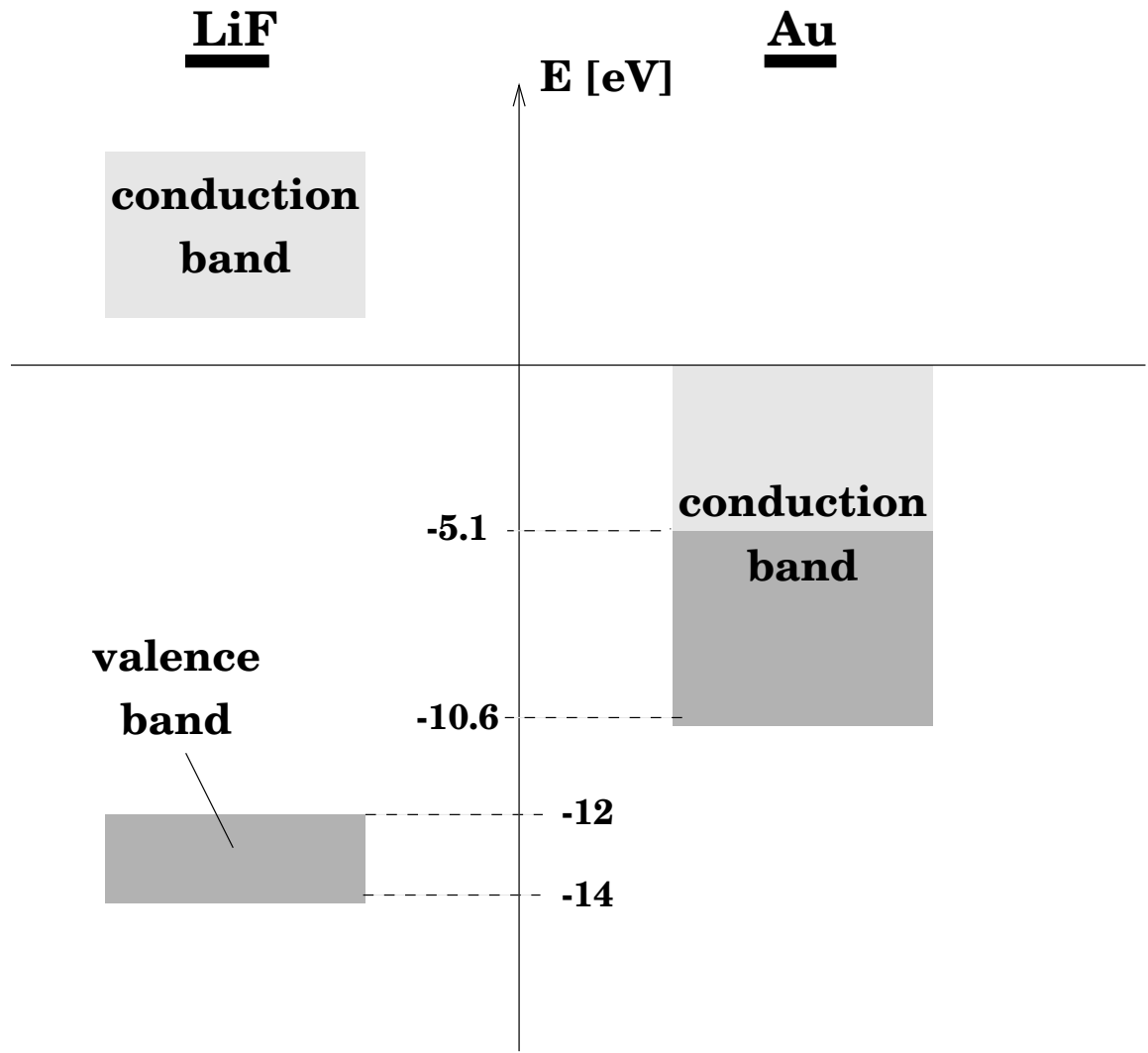
\begin{figure}
  \caption{The band structure of the ionic LiF crystal with a
    large workfunction of 12~eV and a wide band gap of 14~eV. 
    The (polycrystalline) gold
    target represents a typical metal with a workfunction of about
    5~eV and a continuum of unoccupied conduction band states above
    the Fermi level.}
  \label{fig:band}
\end{figure}

Most experiments with incident HCI have been performed for conducting
and semiconducting surfaces with typical workfunctions of about 5~eV.
These experiments focused on total electron yields \cite{Aum93},
energy--resolved Auger \cite{And93,Mey93,Lim94a,Gre95} and X--ray
\cite{Bri90,Sch91} spectra, as well as deflection angles \cite{Win96A}, 
and ion--neutralization \cite{Mey95,Win96,Yan96} measurements.
Recently, several experiments have been carried out, where HCI beams
are incident on insulating surfaces, primarily on ionic crystals, such
as LiF \cite{Aut95,Lim95,Aut96,Lim96}. The unique band structure of
LiF with a large workfunction of 12~eV and a wide band gap of 14~eV
(Fig.~\ref{fig:band}) that elevates the antibinding 2p--band above the
vacuum level provides an interesting opportunity to scrutinize
previous theories about the role of the conduction band in the charge
exchange process. In contrast to metal surfaces, LiF and other
insulating surfaces do not provide unoccupied conduction band states
into which resonant loss from excited projectile states might occur,
and pronounced differences are expected in the neutralization dynamics
of HCIs in front of metal and insulating surfaces. Furthermore, the
capture of electrons from an insulator leads to the local accumulation
of positive surface charges that modify the charge transfer dynamics
in comparison with metals. The capture--induced accumulation of
localized charges on insulator surfaces has very recently been
addressed in a few independent theoretical studies
\cite{Bor96A,Hag97,Cas96} and is a central aspect of the extentions to
the COM discussed in this paper.  Measurements on LiF
surfaces~\cite{Lim96} clearly exhibit the expected discrepancies with
respect to conducting targets and can coherently be interpreted by a
retarded onset for electron capture and characteristic deviations in
the succeeding charge transfer sequence.

In this paper, we adopted the basic framework of the COM suggested by
Burgd\"orfer et al.\ for the interaction of slow HCIs with metal
surfaces \cite{Bur93,Bur91}. Different versions of COMs have been
applied to successfully model charge exchange, energy gain, and
trajectory effects in collisions of HCIs with atoms \cite{Bar85,Nie86}
and clusters ( $\mbox{C}_{60}$) \cite{Bar95,Thu95A,Thu95B,Thu97}.
Within the COM, the neutralization dynamics is described by means of
an effective single electron potential which governs the classical
motion of electrons that are going to be either resonantly captured
into hydrogenic projectile levels or resonantly lost to unoccupied
states of a metal target conduction band. For charge transfer to
occur, an active electron must overcome the potential barrier between
projectile nucleus and target surface. Electron transfer becomes
classically possible if the electron initially occupies a state that
lies above this potential barrier and if vacancies exist in the
resonant final state into which the electron transits. As the
projectile moves along a classical trajectory, both projectile and
target levels experience variable level shifts and change their
relative energetic positions with respect to the potential barrier,
the position and height of which also changes as a function of the
projectile--surface distance.

In order to perform simulations involving insulating surfaces we have
extended the original COM \cite{Bur93,Bur91} by modifying the
dielectric response of the surface to the external projectile charge
and by including local surface charges. These local charges are built
up on the insulator surface during the charge transfer sequence and
decay on a time scale that is given by the conductivity of the
insulator. The local surface charges influence active electrons and
the projectile motion. In this work we will discuss the influence of
these excess surface charges on electron transfer and projectile
deflection in detail and comment on the recent and related theoretical
work of Borisov et al. \cite{Bor96A} and H\"agg et al. \cite{Hag97}.

We have organized this paper as follows. In Section~\ref{sec:outline}
we review the main physical elements of our COM simulations, such as
effective potentials (Sec.~\ref{sec:COM}), local workfunction changes
(Sec.~\ref{sec:Extension}), electronic transition rates
(Sec.~\ref{sec:Transitionrates}), and the projectile motion
(Sec.~\ref{sec:Projmotion}). In Section~\ref{sec:dynamics} we discuss
in detail the neutralization dynamics in front of the surface in terms
of the evolution of level occupations, potentials, projectile charge
and motion above metal and insulator surfaces.
In Section~\ref{sec:image} we compare our results with previously
published experimental and computed data on image energy gains of the
projectile over a wide range of initial charge states.
%
Our conclusions are contained in Section~\ref{sec:conclusions}. We use
atomic units (a.u.) throughout this paper unless specified otherwise.

\section{Outline of the extended COM}
\label{sec:outline}

\subsection{Potentials seen by an active electron}
\label{sec:COM}
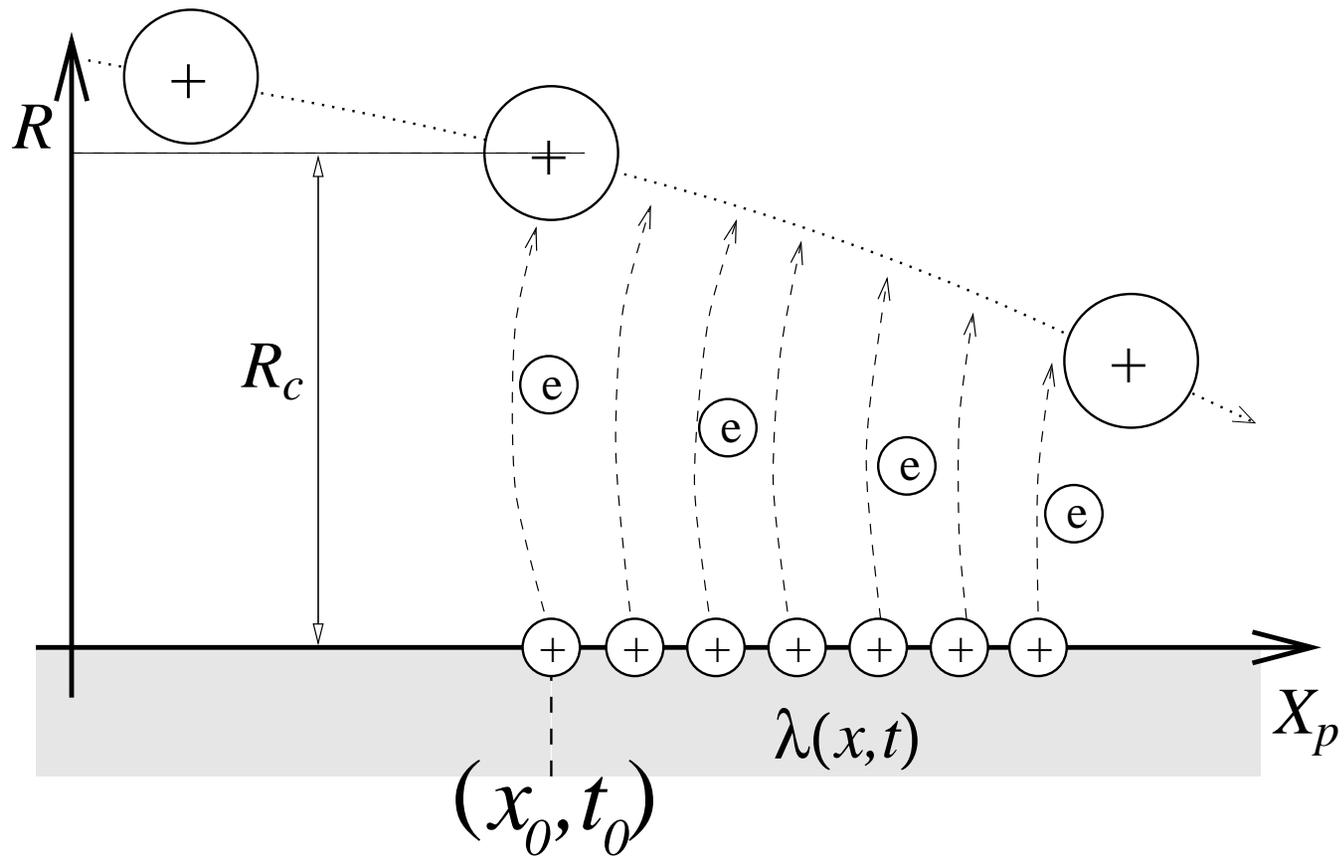
\begin{figure}
  \caption{The linear capture--induced charge distribution $\lambda(x,t)$
    trailing the path of the ion (schematically). The capture sequence
    starts when the incident highly charged ion reaches the critical
    over--barrier distance at a position $(X_p,R) = (x_0, R_c)$ at
    time $t_0$. For our applications to ionic crystals, we assume that
    electrons are captured from the closest surface anion.}
  \label{fig:qtail}
\end{figure}

Within the dynamical COM charge exchange is described in terms of
classical charge currents between energetically shifted valence
states of the target and shifted hydrogenic projectile levels.
These continuous charge currents correspond to electronic transition
rates for resonant capture and loss and occur as soon as the potential
barrier $V_b$ of the total effective potential $V_{tot}$ drops below
the target workfunction $W$. The total potential acting on an active
electron is given by
\begin{eqnarray}
  V_{tot}(q_p,x,z,X_p,R,t) &=& V_{proj}(q_p,x-X_p,z-R) +
  V_{im,p}(q_p,x-X_p,z+R)
  \nonumber \\
  &+& V_{im,e}(z) + V_{local}(x,z,t) 
  \label{equ:Vtot}
\end{eqnarray}
where $x$ and $z$ will denote the electronic coordinates in the
collision plane parallel and perpendicular to the surface--projected
motion, respectively. The projectile distance from the surface is
denoted by $R$. The projectile coordinate along the projection of the
trajectory on the surface is $X_p$. The coordinate $x_0 <0$ refers to
the location on the surface where, at time $t_0$, charge transfer
starts (Fig.~\ref{fig:qtail}). The origin of our coordinate system is
located on the intersection of the topmost lattice plane (at $z=R=0$)
and the collision plane. The jellium edge is located half a lattice
constant above the uppermost lattice plane of the crystal.  The
potential saddle is located at $z_b$. The projectile is assumed to
reach its point of closest approach to the surface at time $t=0$, and
the surface projection of this point defines $x = X_p =0$. For our
applications in this paper, it is sufficient to consider trajectories
with surface projections along the [100] direction that
intersect surface lattice points and define the projectile coordinate
$Y_p = 0$ (see Section IIIB2 below). A more general approach would
average over many trajectories with $Y_p$ coordinates inside a surface
unit cell.

The first term in (\ref{equ:Vtot}) represents the interaction of the
active electron with the projectile and is modeled by the Coulomb
potential
\begin{equation}
  \label{equ:Vproj}
  V_{proj}(q_p,x-X_p,z-R) = -\frac{q_p(R)}{ \sqrt{(x-X_p)^2 +
      (z-R)^2} }.
  \nonumber \\
\end{equation}
The projectile charge
\begin{equation}
  q_p(R) = q_{nuc,p} - \sum_{n} a_{n}(R)
\end{equation}
depends on the nuclear charge $q_{nuc,p}$ of the HCI and the
projectile shell occupations $a_n$. The index $n$ labels the principal
quantum number of projectile shells.

Surface charge distributions produced in response to the external
charges of the projectile and the active electron are included in
(\ref{equ:Vtot}) in form of the projectile image potential $ V_{im,p}
$ and the self--image potential of the active electron $ V_{im,e}$.
These potentials can be derived within linear response theory.  Along
an axis that is perpendicular to the surface and includes the
projectile nucleus (i.e.\ for $x=X_p$), approximate expressions for
these potentials, appropriate for grazing--incidence collisions, are
given by~\cite{Ech92}
\begin{equation}
  \label{equ:Vimp}
   V_{im,p}(q_p,0,z+R)=  
  \frac{2 q_p}{\pi v_p} \int_0^\infty d\omega \:
  \mbox{Re} \left( \frac{1 - \epsilon(\omega)}{1 + \epsilon(\omega)} \right
)
  \cdot K_0 \left( \frac{\omega}{v_p} (z+R-2 z_{im}) \right)
\end{equation}
and the self--image potential of the active electron
\begin{equation}
  V_{im,e}(z) = -\frac{1}{\pi v_p} \int_0^\infty d\omega \:
  \mbox{Re} \left( \frac{1 - \epsilon(\omega)}{1 + \epsilon(\omega)} \right
)
  \cdot K_0 \left( 2 \frac{\omega}{v_p}  (z-z_{im}) \right)
  \label{equ:Vime}
\end{equation}
where $v_p$ denotes the projectile velocity and $K_0$ is a modified
Bessel function.

Equations (\ref{equ:Vimp}) and (\ref{equ:Vime}) refer to the
dielectric response of the target material to a moving external charge
in the undispersive approximation \cite{Ech92}, for which the
dielectric function $\epsilon(\vec k, \omega)$ is independent of the
momentum $\vec k$. Following reference \cite{Low69}, we approximate
the dielectric function
\begin{equation}
  \label{equ:epsilon}
  \epsilon(\omega) = \epsilon_\infty +
  \frac{\epsilon_0-\epsilon_\infty}
       {1-(\omega/\omega_0)^2-i(\omega/\omega_0) \gamma}
\end{equation}
by its static ($\epsilon_0 = \epsilon(0)$) and optical
($\epsilon_\infty = \epsilon(\infty)$) limits, the resonance frequency
$\omega_0$, and a positive infinitesimal constant $\gamma$.  Table I
contains these constants for the two ionic crystals used in this work.
Both image potentials are referred to the image plane at $z_{im}>0$.
In applications to metals, we identify the image plane with the
jellium edge, such that $z_{im}$ becomes equal to half a lattice
constant.  For ionic insulator crystal targets, we inserted the
negative ion radius for $z_{im}$, i.e.\ we assume that the induced
positive image charge is located in the vicinity of the high density
anionic electron cloud above the target.
A similar independent study of the dynamic dielectric 
response of alkali halides
was recently published by H\"agg et al. \cite{Hag97} in which, as in
(\ref{equ:Vimp}), (\ref{equ:Vime}), and (\ref{equ:epsilon}), the
linear response theory result of Abajo and Echenique \cite{Ech92} was
combined with a non--dispersive single--pole fit to the dielectric
function. The main difference to our approach appears to be the use of
a small but finite damping constant
$\gamma$ in the work of H\"agg et al.. We also note that
B\'ar\'any and Setterlind \cite{Bar95} have developed a COM for
capture from a dielectric sphere of radius $a$ and frequency
independent, non--dispersive dielectric constant $\epsilon$. Their
limit $a \rightarrow \infty$ corresponds to an insulating surface with
simplified image interactions that include dielectric screening
effects in terms of a frequency independent, multiplicative factor $(1
- \epsilon) / (1 + \epsilon)$.

For metal surfaces, we can take the limits $\epsilon \mapsto \infty$
in (\ref{equ:Vimp}) and (\ref{equ:Vime}) and obtain the simple
asymptotic forms \cite{Jac75}
\begin{eqnarray}
  V_{im,p}^{metal}(q_p,0,z+R)  = \frac{q_p}{z+R-2 z_{im}}
\\
 V_{im,e}^{metal}(z) =  -\frac{1}{4 (z-z_{im})}.
\end{eqnarray}
In order to avoid the unphysical singularity of the electron
self--image potential at $z=z_{im}$, we truncate and steadily connect
the total image potential to the bulk potential given by the lower
limit of the metal conduction band by extending the constant bulk
potential to a small distance outside $z_{im}$. Our choice for
$z_{im}$ is a little larger than corresponding values obtained by
fitting LDA calculations \cite{Jen88}. It is, however, sufficiently
realistic within the overall precision of our model.

The capture of electrons from a solid leads to a linear surface charge
distribution along the surface--projected path of the projectile
(Fig.~\ref{fig:qtail}). For a metal surface, these local charge
densities vanish instantaneously and do not influence the charge
exchange sequence or the motion of the projectile due to high surface
plasmon frequencies of the order of $10^{16} \mbox{sec}^{-1}$. In
contrast to metal surfaces, typical decay times for excess surface
charges on insulators are by far too long to compensate local charge
accumulations at the collision time scale. As a consequence the
ion is followed by a linearly stretched trail of surface charges
$q_i(x_i)$ generated at times $t_i, \; i=0,\ldots,i_{max}(X_p)$ at
locations $x_i$, for which the projectile is at distances $R(t_i)$ above
the surface. The charge depends on the ion--surface distance
$R$ through the implicit dependence of $x_i$ on $R$.

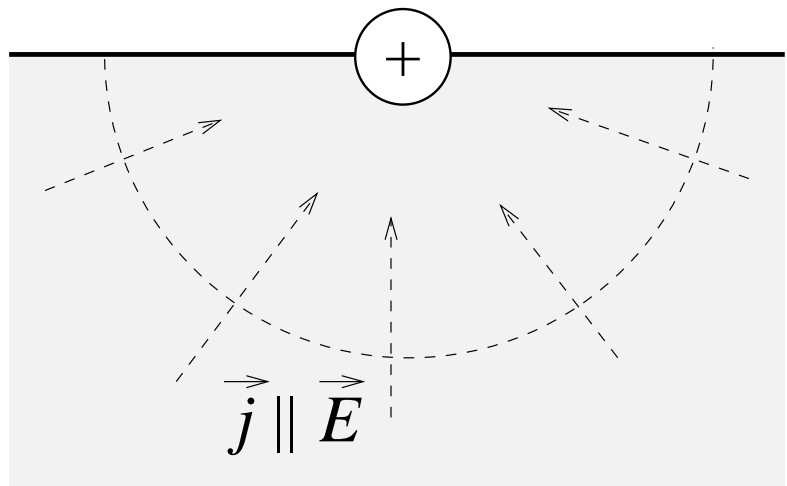
\begin{figure}
  \caption{Classical currents $\vec{j}$ that are driven by the
    field $\vec{E}=\sigma \vec{j}$ of a capture--induced surface
    charge restore electric neutrality. A decay time constant $\tau$
    can be derived from the macroscopic conductivity $\sigma$ by
    applying Gauss' theorem to the current $\vec{j}$ and a Gaussian
    surface given by a hemisphere with the positive excess surface
    charge in its center.}
  \label{fig:surfq}
\end{figure}

We assume the excess surface charges to decay exponentially with a
time constant $\tau$. We can approximate $\tau$ by having charge
currents $\vec{j}$ restore the electric neutrality as depicted in
Figure~\ref{fig:surfq}. The driving force of these currents are
electric fields $\vec{E}=\sigma \vec{j}$ caused by the local surface
charges
\begin{equation}
 \label{equ:qtau}
q_i(x_i,t)=q_i(x_i) \exp \left( -\frac{t-t_i}{\tau} \right), 
                         \;\;\; t\geq t_i
\end{equation}
due to local charges $q_{i}(x_i)$ generated at times $t_i, \;
i=0,1,2,\ldots$ when charge transfer took place.  By choosing a
hemisphere around the excess surface charge as Gaussian surface,
Gauss' theorem leads to the decay time $\tau = (2 \pi \sigma)^{-1}$ in
terms of the macroscopic conductivity $\sigma$. Typical values for
$\sigma$ are $ 10^7 (\Omega \mbox{cm})^{-1}$ for metals and $\sigma
\simeq 1 \cdot 10^{-6} (\Omega \mbox{cm})^{-1}$ for ionic crystals,
such as a LiF crystal, at room temperature. For LiF this simple
estimate yields $\tau \simeq 10^{-7}$s which lies about seven orders
of magnitude above the collision time of typically $10^{-14}$s.  This
means that after capture sets in and while the HCI continues to
interact with the surface, a positive linear charge distribution
\begin{eqnarray}
  \lambda(x,t)&=& \sum_i^{i_{max}(X_p)} q_i(x_i) \exp[-2 \pi \sigma
  (t-t_i)]
  \delta(x - x_i) \nonumber \\
  &\approx& \sum_i^{i_{max}(X_p)} q_i(x_i) \delta(x -x_i) \equiv
  \lambda(x)
  \label{equ:surfcharge}
\end{eqnarray}
remains on the surface--projected projectile path on the ionic
crystal's surface. This charge distribution pulls down the potential
barrier and tends to repel the HCI from the surface. Its contribution
to the total effective potential (\ref{equ:Vtot}) amounts to
\begin{equation}
  \label{equ:Vlocal}
  V_{local}(x,z,t) = - \frac{\epsilon(0) - 1}{\epsilon(0) + 1} \left \{
      \frac{[q_{j,cell}]}{z} +
                     \int_{x_0}^{X_p(t)} dx'
          \frac{\lambda(x',t)}{\sqrt{z^2 + (x' - x)^2}} \right \}.
\end{equation}
The term $[q_{j,cell}]$ represents the integer charge withdrawn from
the active surface cell by an active electron that crosses the barrier, 
e.g.\ $[q_{j,cell}]=1$ for $0<q_{j,cell}\leq 1$.  The
integral in (\ref{equ:Vlocal}) constitutes an average over previously
transferred (non--integer) charges. For metal surfaces, both $\lambda
$ and $V_{local}$ vanish.

The positive excess charge $\lambda(x') dx'$ within a small interval
$dx'$ near $x'$ polarizes the surrounding ionic crystal.  This
polarization effectively screens the local charge $\lambda(x') dx'$
and is the origin of the Mott--Littleton correction \cite{Mot38,Mah80}
to the ground--state energy of the crystal. This screening correction
is approximately included in (\ref{equ:Vlocal}) in terms of the static
dielectric screening function $(\epsilon(0) - 1)/(\epsilon(0) + 1)$. A
Mott--Littleton correction is also included in the work of Borisov et
al. \cite{Bor96A} and H\"agg et al. \cite{Hag97}. Instead of our
multiplicative screening factor in (\ref{equ:Vlocal}) Borisov et al.
introduce this correction as an additive contribution to the
attractive interaction of the active electron with the left--behind
hole on the surface. The approach of H\"agg et al. resembles our
approximation in that the screening of the capture--induced surface
charge is included as a multiplicative factor which asymptotically,
for $z \rightarrow \infty$, becomes equal to a frequency independent
dielectric screening function; it differs from (\ref{equ:Vlocal}) due
to the inclusion of (i) {\it dynamical} screening of excess surface
charges and (ii) fractional ionicity effects close to the surface in
reference \cite{Hag97}.

We note that in our version of the COM a {\it continuous classical\/}
charge current is used to represent charge transfer.  In our
discussion of local surface charges, the discretized charge $q_i$ is
used for convenience only and corresponds to the (small) portion of an
elementary charge that is transferred during one timestep of the
numerical propagation (see Section~\ref{sec:rateqs}, below). In
contrast, Borisov et al. \cite{Bor96A} and H\"agg et al. \cite{Hag97}
enforce charge quantization and consider local excess surface charges
of at least one positive elementary charge.

\subsection{Local workfunction }
\label{sec:Extension}

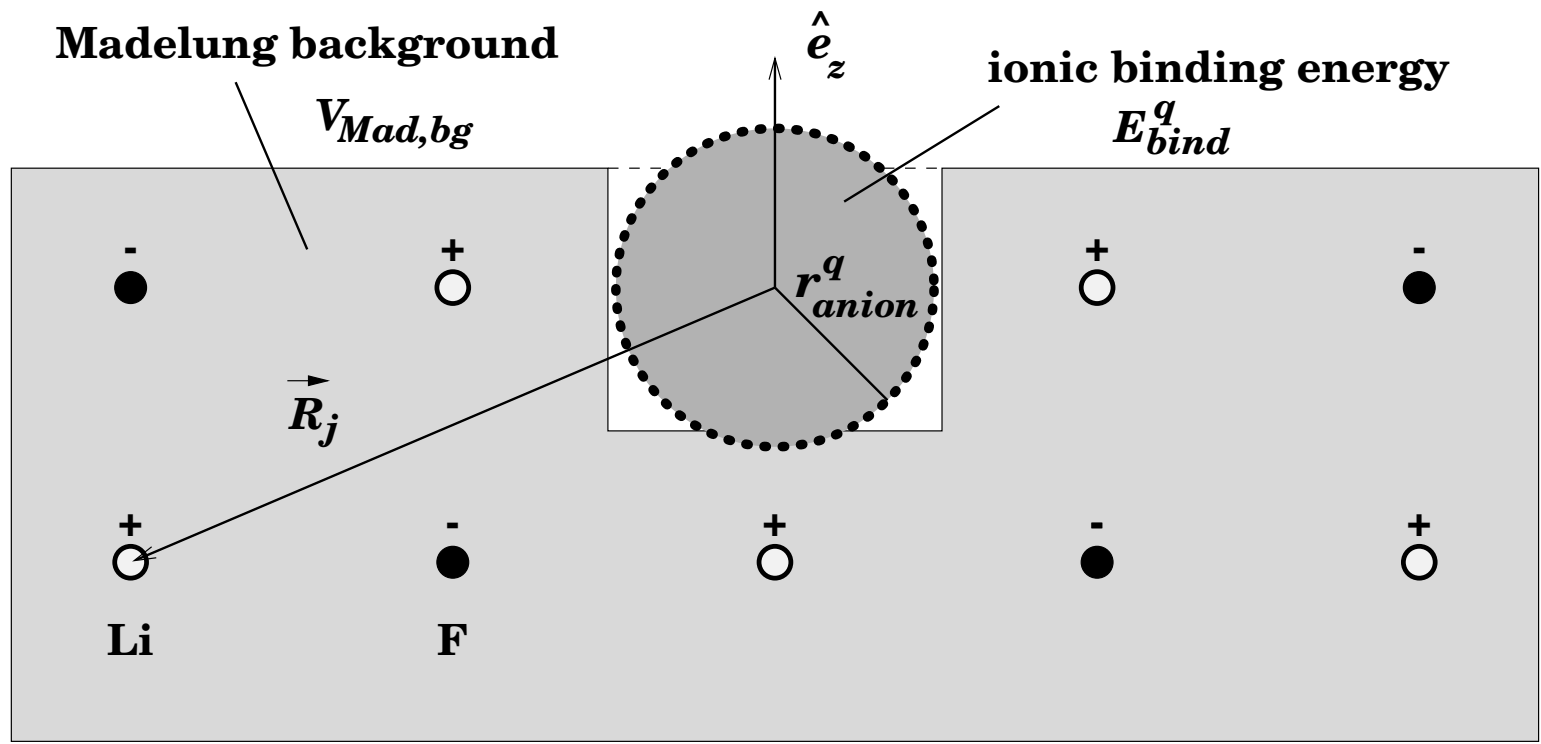
\begin{figure}
  \caption{The charge state dependent local workfunction $W$ of LiF is
    approximated by splitting potentials acting on a bulk atom into an
    anionic contribution, the ionic binding energy $E^q_{bind}$, and
    the Madelung background potential $V_{Mad,bg}$ representing the
    rest of the bulk. The distance $r^q_{anion}$ denotes the orbital
    radius of the most loosely bound subshell of the ionic state and
    is dynamically adjusted to the excess local capture--induced
    charge $q$.}
  \label{fig:madelung}
\end{figure}

We now examine the dynamic change of the local workfunction while the
projectile draws a certain amount of charge from a specific surface
atom on an insulating surface. In order to estimate the local
workfunction of an ionic crystal, we assume that target electrons are
captured from an anion on the surface lattice. The energy necessary to
remove a loosely bound valence electron from a surface anion can be
approximated by the affinity $E^q_{bind}$  of the free anion
(3.4~eV for free F$^-$ ions) and by adding the
interactions of the detaching electron with all other target ions
as a Madelung--background potential $V_{Mad,bg}$
(Fig.~\ref{fig:madelung}). This leads to the workfunction
\begin{equation}
  \label{equ:Wfsplit}
  W(r^q_{anion}) = E^q_{bind} + V_{Mad,bg}(r^q_{anion}).
\end{equation}
We evaluated $V_{Mad,bg}$ at the mean radius $r^q_{anion}$ of
the outermost ionic $n\ell$--shell ($n=2$,$\ell=1$ for LiF), directly
above the anion's lattice site,
\begin{equation}
  \label{equ:Vmad}
  V_{Mad,bg}(r^q_{anion})= \sum_{j}
          \frac{Q_j}{|\vec{R_j} - r^q_{anion}\hat{e}_z |}
\end{equation}
where $ \hat{e}_z$ is a unit vector along the positive z axis.  Since
the contribution of the active surface anion is included via its
binding energy, the active anion is exempted in the Madelung sum over
all lattice sites $\vec R_j$ in (\ref{equ:Vmad}). Without taking
screening effects into account, we assume $Q_j = -1$ for all anionic
charges and $Q_j = 1$ for all cathionic charges in (\ref{equ:Vmad}).
Equation (\ref{equ:Wfsplit}) yields values of 11.65~eV for LiF and
8.21~eV for KI and thus reproduces the experimentally determined
workfunctions (12~eV for LiF and 8.2~eV for KI \cite{Pia81})
sufficiently well within the overall accuracy of our approach.

In order to include the effect of the net capture--induced local
charge $q_{j,cell}>0$ residing within the active surface lattice cell
$j$ on the local workfunction, we added $q_{j,cell}$ to the original
anionic charge $Q_j=-1$ and obtain the new charge $q=Q_j+q_{j,cell}$,
a corresponding new ionic binding energy $E^{q}_{bind}$ and, by using
the Cowan code\cite{Cow81}, an adjusted ionic radius $r^q_{anion}$.
Inserting these quantities into both terms of (\ref{equ:Wfsplit})
supplies the adjusted workfunction $W(r^q_{anion})$. Since the
dynamical COM simulates continuous charge currents, we interpolate
between discrete $r^q_{anion}$--values.
In this way we can compute the local workfunction (\ref{equ:Wfsplit})
as a function of the capture--induced local surface charge
$q_{j,cell}$ within the active anion's unit cell. In contrast to the
approach of H\"agg et al. \cite{Hag97}, we do not include fractional
ionicity and screening effects in the local workfunction. In agreement
with the papers of Borisov et al. \cite{Bor96A} and H\"agg et al.
\cite{Hag97}, our local workfunction includes a Madelung sum for the
interaction of the detaching/ionizing negative charge 
with the ionic crystal and an
additional term for the interaction with the excess surface charge on
the active anion site. However, in contrast to these authors, we
relate the additional term to the affinity of a {\it free} anion of
charge $Q_j = -1$ and interpolate (using atomic ionization potentials)
to effective charges $q > -1$, as dictated by the
non--charge--quantized version of the COM.  For LiF, we allow for at
most one electron to be captured from an active F$^-$ site, such that
$q_{j,cell} \leq 1 $.  For KI, we take the large number of outer shell
electrons on I$^-$ in account by removing this restriction.

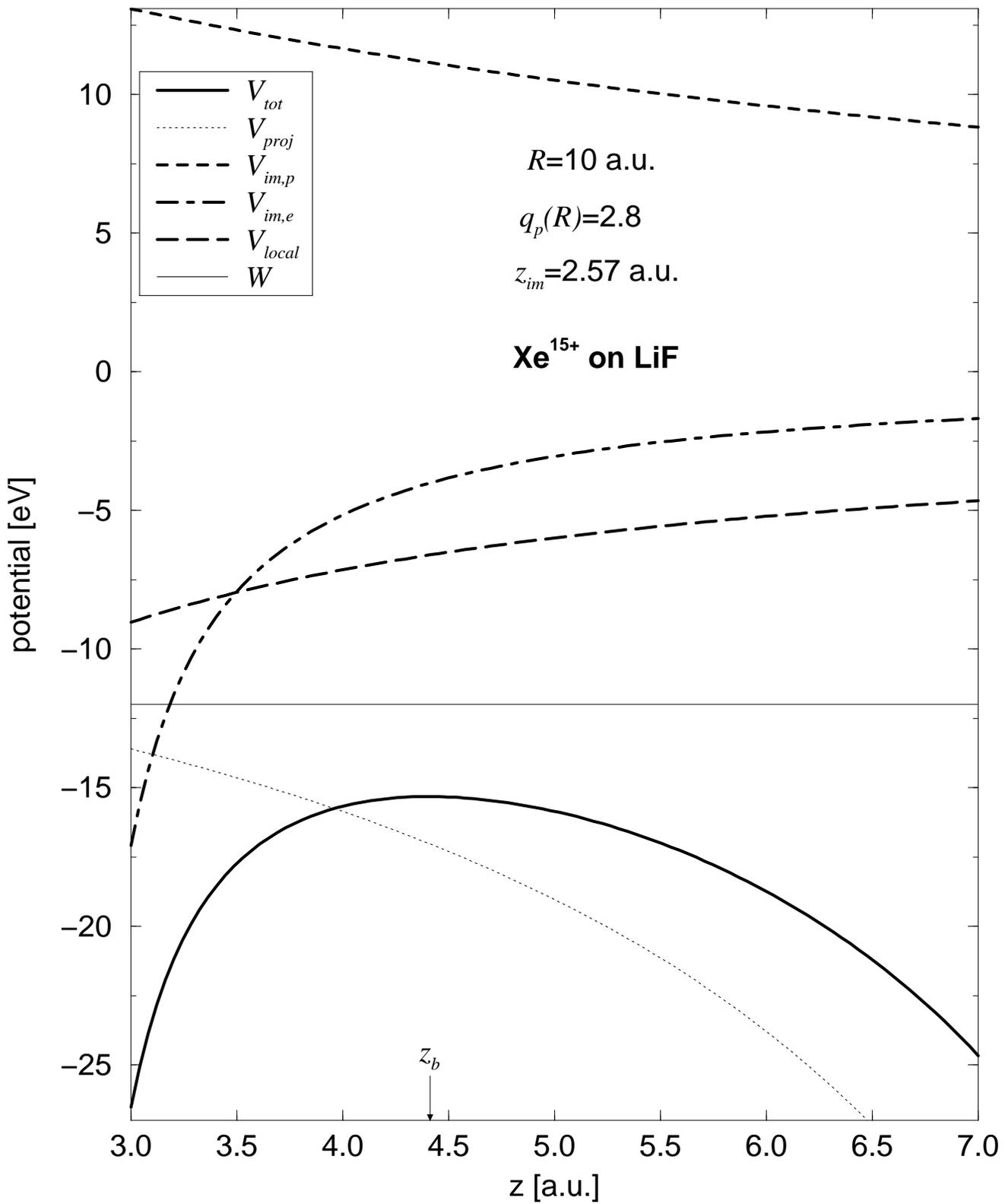
\begin{figure}
 \caption{ Contributions to the total electronic potential $V_{tot}$
   for 50~keV $\mbox{Xe}^{15+}$ approaching a LiF surface at an angle
   of 1$^\circ$ along an axis perpendicular to the surface that
   includes the projectile nucleus. The projectile is on the incident
   part of the trajectory at $R=10$ and $q_p(R=10)=2.8$.}
  \label{fig:potentials}
\end{figure}

In Figure~\ref{fig:potentials} we show the various contributions to
the total electronic potential (\ref{equ:Vtot}) for 50~keV
$\mbox{Xe}^{15+}$ ions that approach a LiF surface at a grazing angle of
1$^\circ$.  The potentials are displayed along an axis perpendicular
to the surface that includes the projectile nucleus. The projectile is
on the incident part of the trajectory at a distance $R=10$ a.u.\ in
front of the surface. The remaining projectile charge $q_p(R=10)$
amounts to 2.8, i.e.\ $\int_{x_0}^{X_p} \lambda(x´)dx´=12.2$.

\subsection{Transition rates}
\label{sec:Transitionrates}

In this section, we summarize the approximations that lead to simple
analytical expressions for resonant and Auger transition rates,
closely following references \cite{Bur93,Bur91}. These rates are then
combined in form of a system of coupled classical rate equations in
order to describe the occupation dynamics of projectile levels.

\subsubsection{Resonant gain}
\label{sec:resgain}

We represent the electronic structure of the projectile by its
spectrum of energy levels $\varepsilon_n(R)$ and their occupations
$a_n(R)$. Both quantities change during the motion of the projectile.
We assume hydrogenic shells with binding energies
\begin{equation}
  \label{equ:levels}
  \varepsilon_n(R) = - \frac{1}{2} \left( \frac{q_{eff,n}(R)}{n}
  \right)^2
\end{equation}
that depend on the effective charges
\begin{equation}
  q_{eff,n}(R) = q_{nuc,p} - \sum_{n'} S_{n,n'} a_{n'}(R).
\end{equation}
The matrix $S_{n,n'}$ accounts for screening effects and is determined
under the simplifying assumption of full inner screening and no
screening by outer and equivalent electrons ($S_{n,n'}=1$ for $n>n'$
and $S_{n,n'}=0$ otherwise). The classical model of a continuous
charge flow over the potential barrier in conjunction with discrete
energy levels (\ref{equ:levels}) requires the definition of energy
bins. We designate energy bins by $[\varepsilon_n(R)]$. Each bin is
attributed to a projectile shell $n$ and includes electronic energies
$\varepsilon$ within the interval
\begin{equation}
  \label{equ:bins}
 [\varepsilon_n(R)] \equiv
 [\frac{1}{2} (\varepsilon_n - \varepsilon_{n-1})
      + V_{im,p} + V_{im,e} + V_{local} \; , \;
  \frac{1}{2} (\varepsilon_{n+1} - \varepsilon_n)
      + V_{im,p} + V_{im,e} + V_{local}] 
\end{equation}
including corrections for
level shifts due to image charge interactions and
localized surface charges. Thus the charge current $I$ transferred
classically from the surface into the energy bin $[\varepsilon_n(R)]$
is considered to feed the $n$'th energetically shifted shell of the
projectile.

The resonant gain current (i.e.\ the resonant gain rate) $I_n^{rg}$
from the surface into a particular $n$-manifold of the HCI is given by
the product
\begin{equation}
  \label{equ:I}
  I_n^{rg}(R,t) = \sigma(R) j_n(R,t)
\end{equation}
of the current density
\begin{equation}
  \label{equ:j}
  j_n(z) = \frac{1}{4}
  \int_{max(V_b,\varepsilon_{n-1/2})}^{min(-W,\varepsilon_{n+1/2})}
  dE D(E) \sqrt{2 (E - V_b)},
\end{equation}
and the cross section $\sigma$. The conduction band density of states
in free--electron--gas approximation is given by
$D(E)={\it V} \sqrt{2}/\pi^2\sqrt{E - V_0}$, where $-V_0$ is the lower
valence band limit with respect to the ionization threshold (cf.
Fig.~\ref{fig:band}). Values for the targets investigated in this work
are $V_0=10.6$, $10.9$, $15.9$, $14$ and $10.5$~eV, respectively for
Au(polycrystalline)~\cite{Lem96,Ash76}, Au(110)~\cite{Lem96,Ash76},
Al~\cite{Ash76}, LiF~\cite{Pia81}, and KI~\cite{OBr74}.  ${\it V}$ is
a volume which we assume to be one (a.u.)$^3$ for the following.  The
factor $\sqrt{2 (E - V_b)}$ is the classical velocity of active
electrons while passing the potential barrier. The energetic bottom of
the conduction band lies $V_0$ below the ionization threshold, and the
geometrical factor $1/4$ in (\ref{equ:j}) relates the isotropic
density of states, $D(E)$, to the electron current along the positive
surface normal.
The cross section
\begin{equation}
 \sigma = \pi \left( \frac{\Delta x}{2} \right)^2
\end{equation}
is equal to the classically allowed area over the potential saddle
through which the current representing active electrons needs to flow.
The effective width $\Delta x(t) = |x_1(t) - x_2(t)|$ of the saddle at
any time is given implicitly by the two solutions $x_1$ and $x_2$ of
\begin{equation}
  \label{equ:width}
  \varepsilon_n(R) + V_{im,p}(q_p,x-X_p,z_b+R) + V_{im,e}(z_b) +
  V_{local}(x,z_b,t) = V_{tot}(q_p,x,z_b,X_p,R,t)
\end{equation}
where, as in (\ref{equ:Vimp}), the nuclear image potential is
evaluated on the axis $x = X_p$.

We note that the valence band density of states of an ionic crystal
per se is poorly represented with the free electron gas model. H\"agg
et al. \cite{Hag97} used classical Monte Carlo techniques in order to
simulate the over--barrier dynamics of target electrons that are
released from an anionic center by the highly charged projectile.
Their Monte Carlo study indicates that the electron is effectively
captured from LiF at a projectile surface distance that is about 3
a.u.\ closer to the surface than the onset of classical electron
capture across the potential barrier. Interestingly, our numerical
results, including capture rates modeled with reference to the free
electron density of states, also indicate a delayed onset between the
initiation of a classical over--barrier current and the
projectile--surface distance where one elementary charge has been
transferred (see Section IIIA below).  Furthermore, we point out that
previous dynamical COM studies on electron capture in collisions with
C$_{60}$ \cite{Thu95A} have shown a rather weak dependence of final
projectile charge states and critical capture radii on variations in
the resonant capture rates. We therefore conclude, that within the
overall accuracy of the COM and in view of the narrow valence band of
ionic crystals, the free electron gas model is sufficiently realistic
for providing acceptable estimates for resonant capture rates.  The
agreement of our simulated projectile energy gains with experiments
for ionic crystal surfaces provides further support for this
approximation (see Section IIIB2, below).

\subsubsection{Resonant loss}
\label{sec:resloss}

The rate of electron loss from atomic energy levels into unoccupied
bulk levels can be obtained from the electron's orbital frequency of
revolution
\begin{equation}
  \label{equ:revolution}
  f_n = \frac{q_{eff,n}^2}{2 \pi n^3}
\end{equation}
multiplied with its probability to hit the saddle region that is
approximately given by
\begin{equation}
  \label{equ:sadreg}
  P(\varepsilon_n) =
  [z_{crit}(\varepsilon_n)-z]/z_{crit}(\varepsilon_n)
\end{equation}
where $z_{crit}(\varepsilon_n)$ denotes the critical distance where
the first electron capture occurs into shell $n$ \cite{Bur91},
\begin{equation}
 I_n^{rl}(z) = f_n \cdot P(\varepsilon_n).
\end{equation}
For insulators, particularly for LiF with no unoccupied band levels
below the vacuum level, resonant loss is irrelevant, and $I_n^{rl}=0$.

\subsubsection{Auger processes}
\label{sec:augerproc}

Intraatomic Auger transitions induce small changes in the projectile
occupation during the interaction with the surface. Following
reference \cite{Bur91} we express the Auger rates by a simple analytic
fit through data points
\begin{equation}
  \label{equ:Augerates}
  \Gamma_{n_i,n_f} = \frac{5.06 \cdot 10^{-3}}{( n_i-n_f)^{3.46}}
\end{equation}
that have been calculated with the Cowan code \cite{Cow81} for fast
transitions between two given shells $n_i$ and $n_f$.

\subsubsection{Rate equations}
\label{sec:rateqs}

The dynamically varying projectile populations are obtained as
solutions to the coupled set of classical rate equations
\begin{equation}
  \label{equ:da/dt}
  \frac{da_n}{dt} = \Theta(A_n-a_n) \Gamma_{n}^{rg} - a_n \Gamma_{n}^{rl}
 +
  w_{f,n} \sum_{n'>n} \Gamma_{n',n} w_{i,n'} -
  2 w_{i,n} \sum_{n'<n} \Gamma_{n,n'} w_{f,n'}
\end{equation}
where the degeneracy of shell $n$ is given by $A_n = 2n^2$. $\Theta$
is the unit step function. The (empirical) statistical factor $w_{f,n}
= 1/(1 + 1.5 a_n)$ corrects for the decrease in Auger transition rates
due to increasing populations $a_n$ of the final level. The
statistical factor $w_{i,n} = \frac{1}{2} a_n (a_n-1)$ takes the
equivalence of electrons in the initial shell into account.

\subsection{Projectile motion}
\label{sec:Projmotion}

Before reaching the first critical over--barrier radius, the motion of
the incoming HCI is solely affected by its attractive self--image
force. For metal surfaces this force is given by
\begin{equation}
\label{equ:Fimpp}
 F_{im,p-p}(R) = - \left(\frac{q_p}{2 (R-z_{im})} \right)^2
\end{equation}
and for insulating surfaces by the derivative of (\ref{equ:Vimp}).
After the projectile has reached the first critical over--barrier
distance, charge transfer begins and, for insulators, the self--image
force (\ref{equ:Fimpp}) starts to compete with the repulsive force
created by localized surface charges (\ref{equ:Vlocal}).  For grazing
collisions, the latter force is weak, due to the large projectile
velocity component parallel to the surface, which rapidly increases
the distance between previously created surface charges and the HCI.

At distances $R$ smaller than the largest radius of occupied atomic
orbitals, $\langle r \rangle_n$, the electron clouds of the incoming
HCI and the surface ions begin to penetrate each other. The accurate
description of this situation would require detailed quantum dynamical
calculations, which are far beyond the overall simplistic nature and
accuracy of the COM. In order to determine the classical motion of the
projectile, we employ the Thomas--Fermi model and use the
Thomas--Fermi--Moli\`ere potential energy \cite{Mol47},
\begin{equation}
  \label{equ:VTFM}
  W_{TFM}(r) = \frac{q_{nuc,p} \cdot q_{nuc,t}}{r} \cdot
  \phi\left(\frac{r}{a}\right).
\end{equation}
This interatomic potential includes the Coulomb repulsion between the
two nuclear charges $q_{nuc,p}$ and $q_{nuc,t}$ of the HCI and a
target atom, respectively, and a screening function $\phi$ that
depends on the internuclear distance $r$ scaled by the screening
length $a$,
\begin{eqnarray}
  \label{equ:phi}
  \phi(\frac{r}{a}) &=& \sum_{i=1}^3 \alpha_i \exp\left(-\beta_i 
  \frac{r}{a}\right)  \\
        a&=& 0.88534 / \sqrt{q_{nuc,p}^{2/3} + q_{nuc,t}^{2/3}} \;, \nonumber
\end{eqnarray}
with $\{\alpha_i\}=\{0.35,0.55,0.10\}, \{\beta_i\}=\{0.3,1.2,6.0\}$.

For for small perpendicular projectile velocity components, as
typically given in grazing incidence collisions, the interaction of
the HCI with an array of surface atoms or ions is well represented by
taking the planar average of (\ref{equ:VTFM}) \cite{Gem74},
\begin{eqnarray}
  \label{equ:VTFMpl}
  W_{TFM}(R) = \frac{2 \pi a q_{nuc,p} q_{nuc,t}}{d^2}
  \phi'\left(\frac{R}{a}\right) \nonumber \\ \phi'(\frac{R}{a}) =
  \sum_{i=1}^3 \frac{\alpha_i}{\beta_i} \exp\left(-\beta_i
    \frac{R}{a}\right)
\end{eqnarray}
where, for simplicity, we have assumed a square lattice 
with lattice constant $d$.  

For ionic crystals that are composed of two different ion species, we
apply (\ref{equ:VTFMpl}) separately to surface lattices of anions and
cathions. This results in the planar averaged potential
\begin{equation}
 \label{equ:VTFMpl1}
 W_{TFM}(R) = W_{TFM}^{anion}(R) + W_{TFM}^{cathion}(R)
\end{equation}
where $ W_{TFM}^{anion} + W_{TFM}^{cathion} $ are constructed
according to (\ref{equ:VTFMpl}) with $q_{nuc,t}$ replaced by the
respective nuclear charges of anions and cathions and with the
distance $d$ between anions or between cathions, respectively.

The force exerted on the projectile by the capture--induced surface
charge distribution (\ref{equ:surfcharge}) is repulsive with a
parallel component that accelerates the projectile in positive
x--direction. It is given by
\begin{equation}
 \label{equ:Flocal}
 \vec F_{local}(X_p,R) = \frac{ \epsilon(0) -1}{ \epsilon(0) +1} q_p
 \int_{x_0}^{X_p} dx' \frac{ \lambda(x',t) \; (R\, ,\, X_p -x')}{ (R^2
   +(X_p -x')^2)^{3/2}}  .
\end{equation}
The factor $(\epsilon(0) -1)/(\epsilon(0) +1)$ accounts for the static
dielectric screening (cf. equ. (\ref{equ:Vlocal})).  The effect of
this force on the projectile motion parallel to the surface is small.
For 30~keV Xe$^{15+}$ incident under a grazing angle of 1$^\circ$ on
LiF, it changes the parallel component of the projectile velocity by
0.2~\%.

The net force on the projectile is now given by the negative gradient
of $W_{TFM}$, the projectile self--image force (\ref{equ:Fimpp}), and
(\ref{equ:Flocal}),
\begin{equation}
 \label{equ:Vpnet}
 \vec F_{net}(X_p,R) = \left\{ -\frac{d}{d R} ( W_{TFM}(R) )
         + F_{im,p-p}(R) \right\} \, \hat{e}_z + \vec F_{local}(X_p,R)   .
\end{equation}

\section{Numerical results and discussion}
\label{sec:discussion}

In this section we are going to discuss the pronounced differences in
the above--surface neutralization dynamics between insulating and
conducting targets as predicted by our simulations.
All electronic interactions depend strongly on the projectile's
position on its trajectory $(X_p(t),R(t))$, which, in turn, depends on
the charge state evolution $q_p(t)$ and, for insulator targets, on the
trail of positive excess surface charges.  Due to this coupling of
nuclear and electronic degrees of freedom, the reconstruction of
measured projectile deflection angles and projectile energy gains will
supply support for the specific interaction model implemented in our
simulation. Even though it is not yet possible to extract direct
evidence from the experimental data for the time evolution of many
quantities included in our simulation, such as shell populations and
level shifts (Sec.~\ref{sec:dynamics}), good agreement between energy
gain measurements and theory (Sec.~\ref{sec:image}) supports
the validity of our model assumptions.

\subsection{Interaction dynamics}
\label{sec:dynamics}

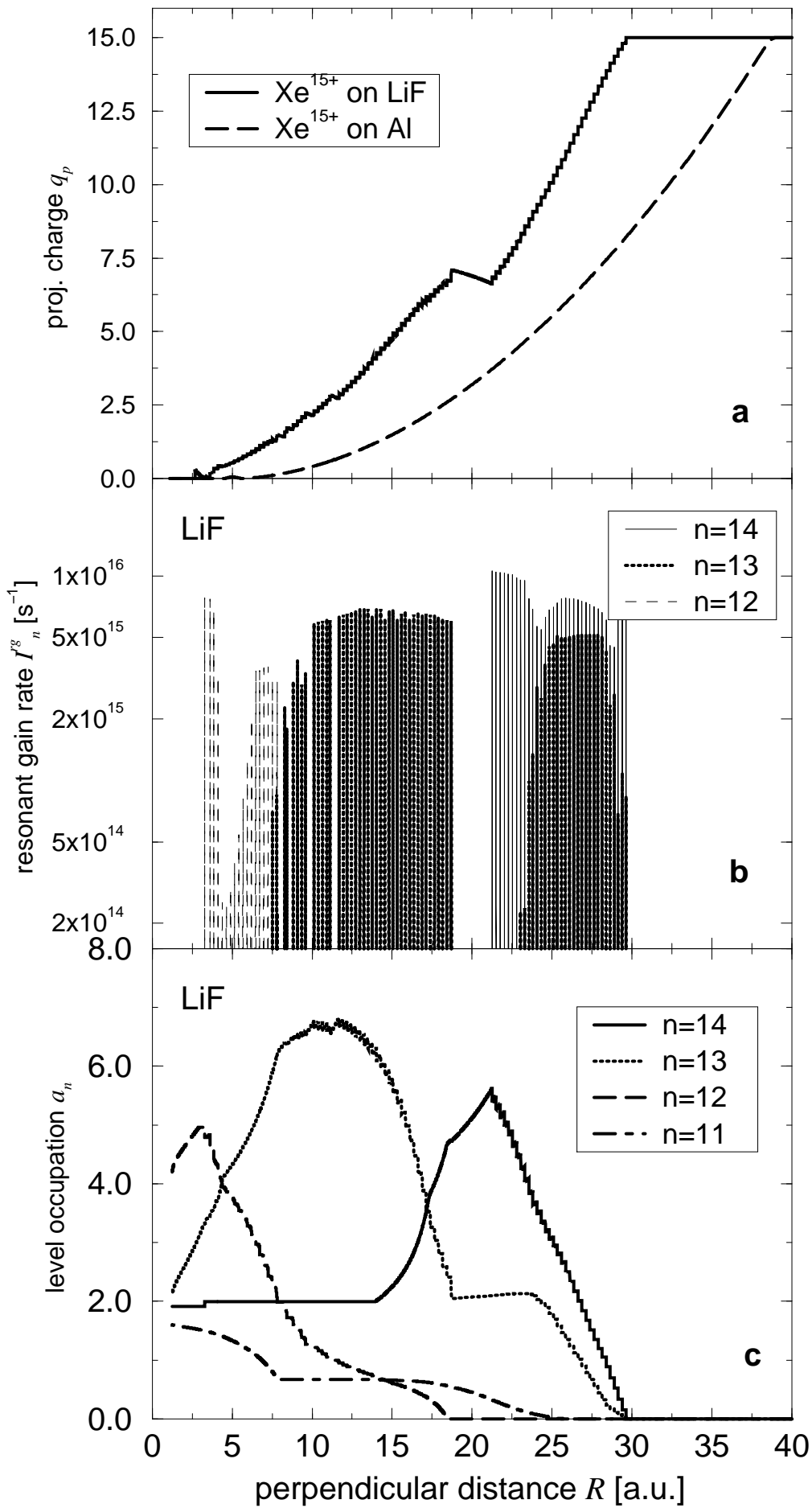
\begin{figure}
  \caption{Results for $\mbox{Xe}^{15+}$ at $E_{kin}=50$~keV and an
    incidence angle of $1^\circ$. Time evolution of the projectile
    charge state for LiF and Al surfaces (a). Resonant gain rates on
    LiF for the two highest resonantly populated shells (b).
    Projectile shell occupations on LiF (c).}
  \label{fig:Xe15LiF:tq:rg:do}
\end{figure}

Fig.~\ref{fig:Xe15LiF:tq:rg:do}a shows the simulated projectile charge
evolution for $\mbox{Xe}^{15+}$ ions colliding with an Al target in
one case, and a LiF crystal in the other. The incident projectile
energy is 50~keV at a grazing incidence angle of $1^\circ$. For Al,
the first critical distance for classical over--barrier capture is
$R_c \approx 38$. In the case of LiF, the interplay of the large
workfunction and image charges, which compared with the Al target are
reduced by the altered dielectric response function, effectively
shifts the onset of charge transfer by about 10 a.u.\ to $R_c \approx
30$. Our version of the COM does not impose charge quantization. By
rounding to nearest integer charges, the neutralization sequence is
therefore completed when the projectile charge becomes smaller than
0.5 . In comparison with LiF, we find that for the same ion--surface
distance $R$ the early onset of electron transfer on Al leads to
smaller projectile charges above the metal target.

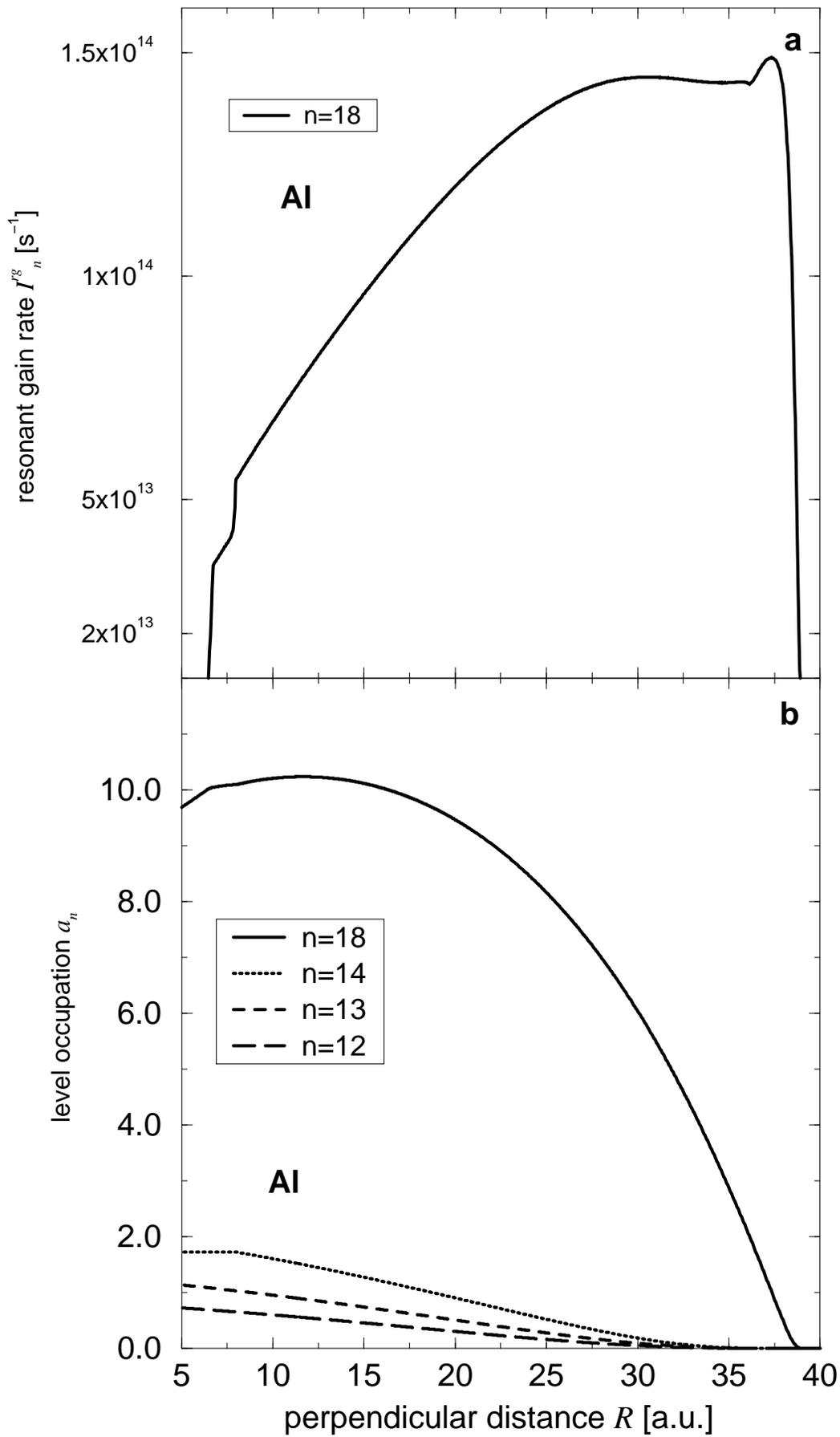
\begin{figure}
  \caption{Results for $\mbox{Xe}^{15+}$ ($E_{kin}=50$~keV) ions
    impinging under $1^\circ$ grazing incidence conditions on an Al
    surface.  Time evolution of the resonant gain rates for the
    highest resonantly populated shells (a).  Shell occupation (b).
    Charge exchange primarily takes place via resonant gain into the
    $n=18$ shell.}
  \label{fig:Xe15Al:rg:rl:do}
\end{figure}

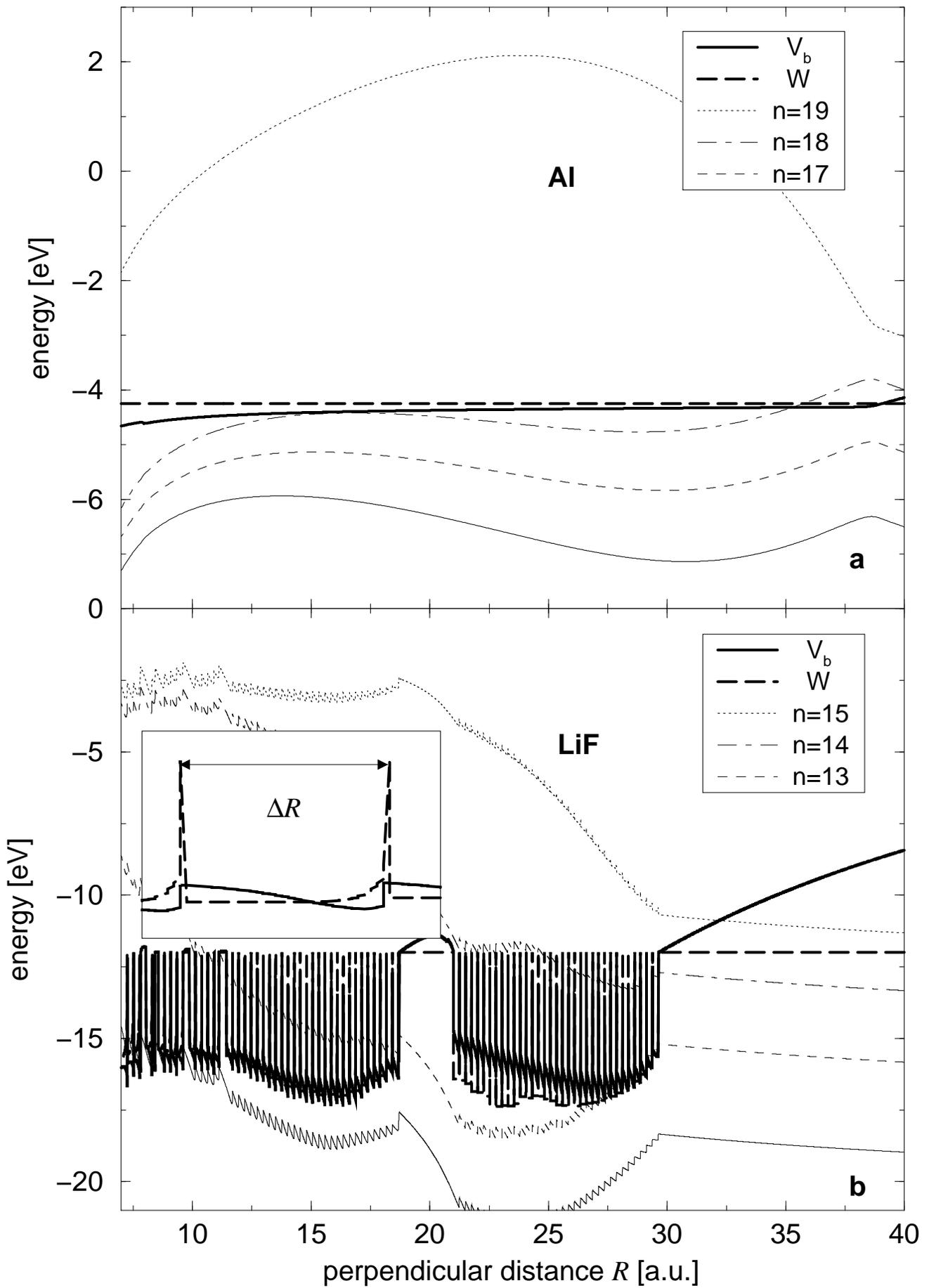
\begin{figure}
  \caption{Results for $\mbox{Xe}^{15+}$ ions impinging with
    $E_{kin}=50$~keV at an grazing angle of $1^\circ$ on Al and LiF
    surfaces. The plot shows the time evolution of the potential
    barrier $V_b$, the target workfunction $W$, and the projectile
    shell occupations of the most active shells.  $V_b$ and $W$
    display characteristic oscillations on LiF, whereas all
    potentials evolve smoothly on Al (see text for details).}
  \label{fig:Xe15LiF:Al:ep:sa}
\end{figure}

Differences in the time evolution of resonant gain processes become
apparent by comparing resonant gain rates
(Fig.~\ref{fig:Xe15LiF:tq:rg:do}b and Fig.~\ref{fig:Xe15Al:rg:rl:do}a)
or projectile level occupations (Fig.~\ref{fig:Xe15LiF:tq:rg:do}c and
Fig.~\ref{fig:Xe15Al:rg:rl:do}b) for the two targets and incident
$\mbox{Xe}^{15+}$ ions.  On the Al target, the n=18 shell of the
projectile becomes strongly populated with filling rates of the order
of $10^{14}\mbox{s}^{-1}$ (Fig.~\ref{fig:Xe15Al:rg:rl:do}). For LiF,
resonant gain transfer occurs into the n=14, 13, and 12 shells. The
sharp maxima of the gain rates on LiF exceed the metal rates by almost
two orders of magnitude. The average neutralization rates, however,
i.e.\ the slopes of the corresponding shell occupations
(Fig.~\ref{fig:Xe15LiF:tq:rg:do}a and
(Fig.~\ref{fig:Xe15Al:rg:rl:do}b), are only slightly higher in LiF.
The interruption of the projectile neutralization between $R=19$ and
$R=21$ coincides with the transient increase of $V_b$ above $W$ in
Fig.~\ref{fig:Xe15LiF:Al:ep:sa}(b). Resonant loss processes are either
forbidden (LiF) or contribute with negligible rates (Al). For the Al
target, shells below the resonantly populated level n=18 are populated
in Auger transitions.

The regularly spaced spikes in the resonant gain rate for
$\mbox{Xe}^{15+}$ impinging on LiF (Fig.~\ref{fig:Xe15LiF:tq:rg:do}b)
originate in the capture--induced local surface charges. As the
surface--projected path of the ion enters a new surface cell
containing a single fluorine F$^-$ ion, the workfunction is reset to
its original value $W=12$~eV, such that the local Fermi level suddenly
moves upwards thereby stimulating over--barrier capture. The
corresponding workfunction changes amount to up to 6~eV
(Fig.~\ref{fig:Xe15LiF:Al:ep:sa}b), whereas oscillations in the
barrier height $V_b$ of the potential saddle remain comparatively
small with an amplitude of less than $1$~eV as the ion travels over
the surface cell boundaries.  This relatively inert behavior of $V_b$
can be explained by the moderate influence of the local surface
charges on the total potential $V_{tot}$ near the saddle position
$z_b$ which is situated typically a few a.u.\ in front of the first
bulk layer (see also Fig.~\ref{fig:potentials}). We note that, if
capture from a given anion proceeded, the local workfunction would
increase by about 10~eV per unit capture--induced surface charge.
However, due to the high transfer rates of up to
$10^{16}\mbox{s}^{-1}$ the continuous current of negative charge is
quickly cut off at the moment when the Fermi level is shifted below
the saddle point $V_b$ ("over--barrier cut-off").  In other words, the
local workfunction change (i.e. the shift of the Fermi level towards
lower energies and below $V_b$) generated by capture--induced surface
charges produces the peaked structures in the resonant gain rates.
For the Al target, the absence of local surface charges results in a
comparatively steady evolution of the gap between $V_b$ and $W$
(Fig.~\ref{fig:Xe15LiF:Al:ep:sa}a) and results in the mostly smooth
development of the dominant resonant gain rates in
Figure~\ref{fig:Xe15Al:rg:rl:do}a.

Considering the characteristic discrepancies in the resonant exchange
mechanisms, it is surprising that the average rate of neutralization
is very similar for both targets. The effects of the low alkali halide
workfunction on the onset of charge exchange, the reduced dielectric
response of the insulator, and local surface charges appear to be
counterproductive. As will be shown below (Sec.~\ref{sec:image}), this
interplay is also related to the strikingly small differences between
the image energy gains of a particular HCI on LiF and Al targets.

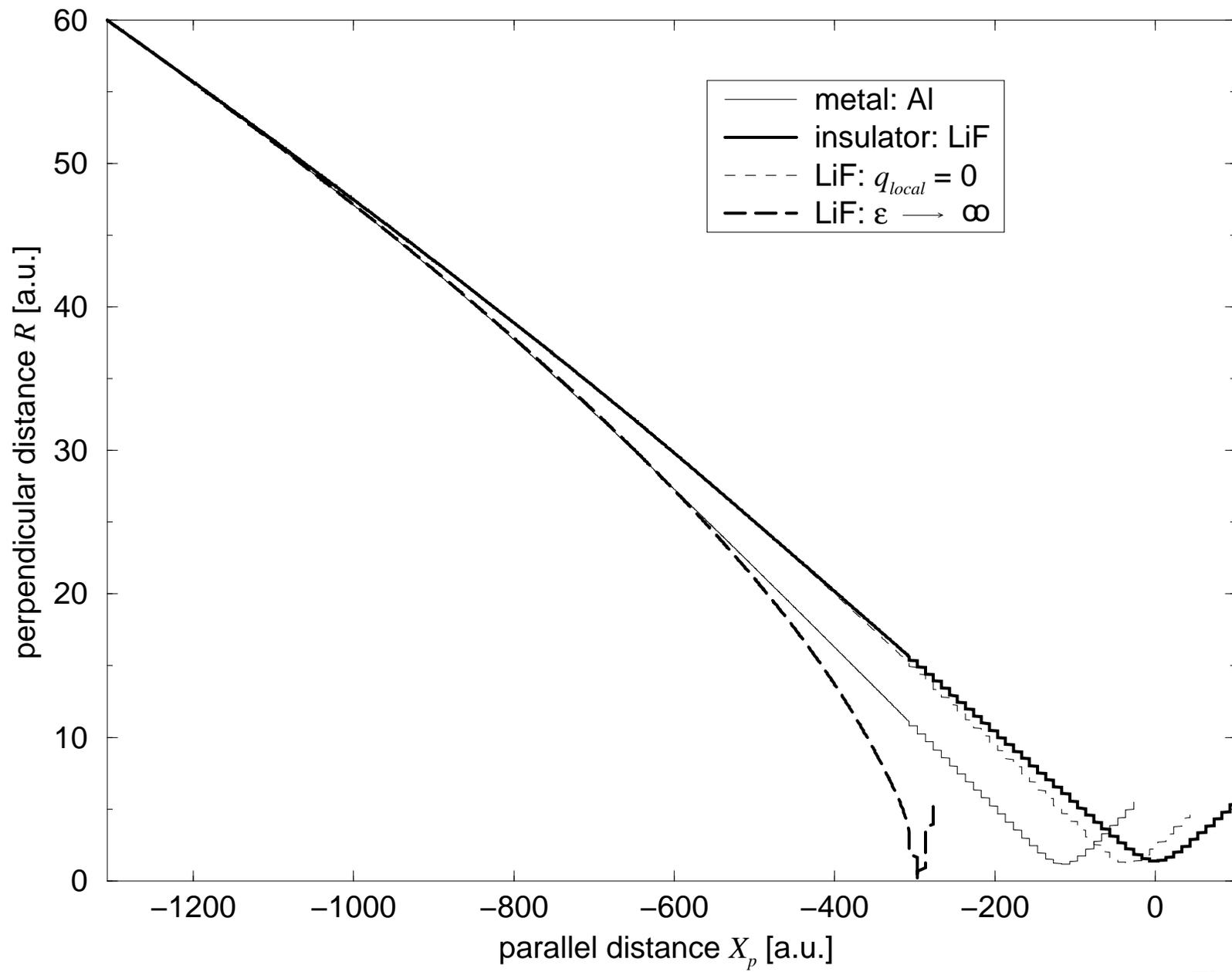
\begin{figure}
  \caption{Projectile trajectory in terms of the perpendicular
    projectile velocity component $v_z$ versus $R$ for
    20~keV--$\mbox{Xe}^{15+}$ ions impinging under $1^\circ$ on Al and
    LiF. The first two curves exhibit standard dynamical COM
    simulations on these targets. The next two curves show simulation
    results for the same $\mbox{Xe}^{15+}$ projectile on LiF when
    local surface charges have been disabled in the third and a large
    ``metal'' value for the dielectric susceptibility $\epsilon
    \mapsto \infty$ has been chosen for the fourth curve. $X_p=0$
    corresponds to the vertex of the full simulation for LiF.}
  \label{fig:Xe15LiF:Al:pt}
\end{figure}

In Figure~\ref{fig:Xe15LiF:Al:pt} we compare ion trajectories for
grazingly incident 20~keV
$\mbox{Xe}^{15+}$ on Al and LiF surfaces. At large
distances, the magnitude of the perpendicular velocity component
$|v_z|$ steadily increases due to the attractive projectile
self--image force. The short range TFM--potential in (\ref{equ:VTFM})
causes the inversion of the trajectory in a small region that measures
about 2 a.u.\ relative to the vertex of the trajectory leading to
nearly specular reflection. For the LiF target the attenuated
dielectric response of the insulator weakens the image attraction in
comparison with a metal target. We simulated this effect in a separate
calculation for LiF where we replaced the insulator specific
dielectric response in the projectile (self--) image interactions by
the asymptotic response of a perfect metal, taking (\ref{equ:Fimpp})
and the limit $\epsilon \rightarrow \infty$ in (\ref{equ:Vimp}). This
yields a noticeable change in the ion trajectory.  Before capture sets
in at large $R$, only the image force acts on the HCI and the ``metal
dielectric response" trajectory for LiF nearly coincides with the ion
trajectory in front of Al. The replacement of the insulator--specific
dielectric response by the metallic response 
on LiF moves the onset of charge transfer
$13$~a.u.\ closer to the surface.  This shifts the potential barrier
upwards due to the more repulsive (unscreened) projectile image term
(\ref{equ:Vimp}), thus reduces the amount of charge captured, and more
than doubles the overall energy gain above the target, clearly leading
away from both our dynamical COM with insulator--specific response and
experimental results (see Sec.~\ref{sec:image}).

In another separate simulation we have eliminated all effects due to
the capture--induced positive surface charge distribution on LiF
(Fig.~\ref{fig:Xe15LiF:Al:pt}). The local surface charges add to the
projectile repulsion near the surface. The vertex is now located about
0.1 a.u.\ closer to the first bulk layer than for the genuine 50~keV
Xe$^{15+}$--LiF simulation including local surface charges. This
small shift suggests that the direct influence
of these surface charges on the projectile trajectory is rather small.
The large discrepancy of $12.8$~eV in the image energy gains between
both simulations (see Section~\ref{sec:insulators}) originates mainly
from deviations in the neutralization dynamics leading to a higher
average projectile charge in front of the surface when local surface
charges are disabled. This can be understood by considering that the
component $V_{local}$ in (\ref{equ:Vlocal}) pulls down the potential
barrier (\ref{equ:Vtot}) and also the projectile energy levels
(\ref{equ:bins}) and thus counteracts electron loss due to level
promotion into the continuum (if an occupied level gets promoted
to the continuum, we assume that the level is instantaneously
ionized).

The distance of closest approach to the surface under grazing
incidence is determined by the initial velocity component
perpendicular to the surface, the total kinetic energy gain of the HCI
at the turning point, and by the composition of the target material
via the screened interatomic interaction (\ref{equ:VTFM}). Our
simulations yield turning points at $R=1.1$ and $1.3$ for
$\mbox{Xe}^{15+}$ ions impinging on Al and LiF, respectively, at an
incident energy of 30~keV and an incidence grazing angle of $1^\circ$.

\subsection{Image energy gains}
\label{sec:image}

After having presented detailed results on the interaction dynamics in
the previous section, we are now going to demonstrate that the
extended dynamical COM can quite accurately reproduce previously
published (measured and simulated) data on image energy gains for both
conducting and insulating crystals over a wide range of initial
projectile charge states. Our simulations as well as recent
experiments \cite{Win96} show that the neutralization of the HCI is
completed prior its reflection for a wide range of initial projectile
charge states. The inversion of the perpendicular velocity component
$v_\perp$ takes place at a distance of a few atomic units above the
surface (Fig.~\ref{fig:Xe15LiF:Al:pt}), and the measurable difference
between the asymptotic incident and reflection angles of an ion beam
can be straightforwardly correlated to the net image energy gain
\cite{Win93,Win96A}.

\subsubsection{Metals}
\label{sec:metals}

Image energy gains of a HCI impinging on metal surfaces are
characterized by an approximate $q_p^{3/2}$--increase with the initial
projectile charge state $q_p$ \cite{Bur96,Lem96,Win96A}. A lower limit
for the energy gain can be deduced by assuming that the projectile is
instantaneously and completely neutralized at the first critical
distance $R_c \simeq \sqrt{8 q_p+2}/(2W)$ \cite{Bur93A}. We shall
refer to this estimate as {\it "simple COM"}. The energy gain for
large $q_p$ is then given by the analytical formula
\begin{equation}
\Delta E = \frac{q_p^2}{4 R_c} = \frac{W q_p^{3/2}}{4 \sqrt{2}}.
\end{equation}
Consequently, $\Delta E / W$ should be independent of the
target material.

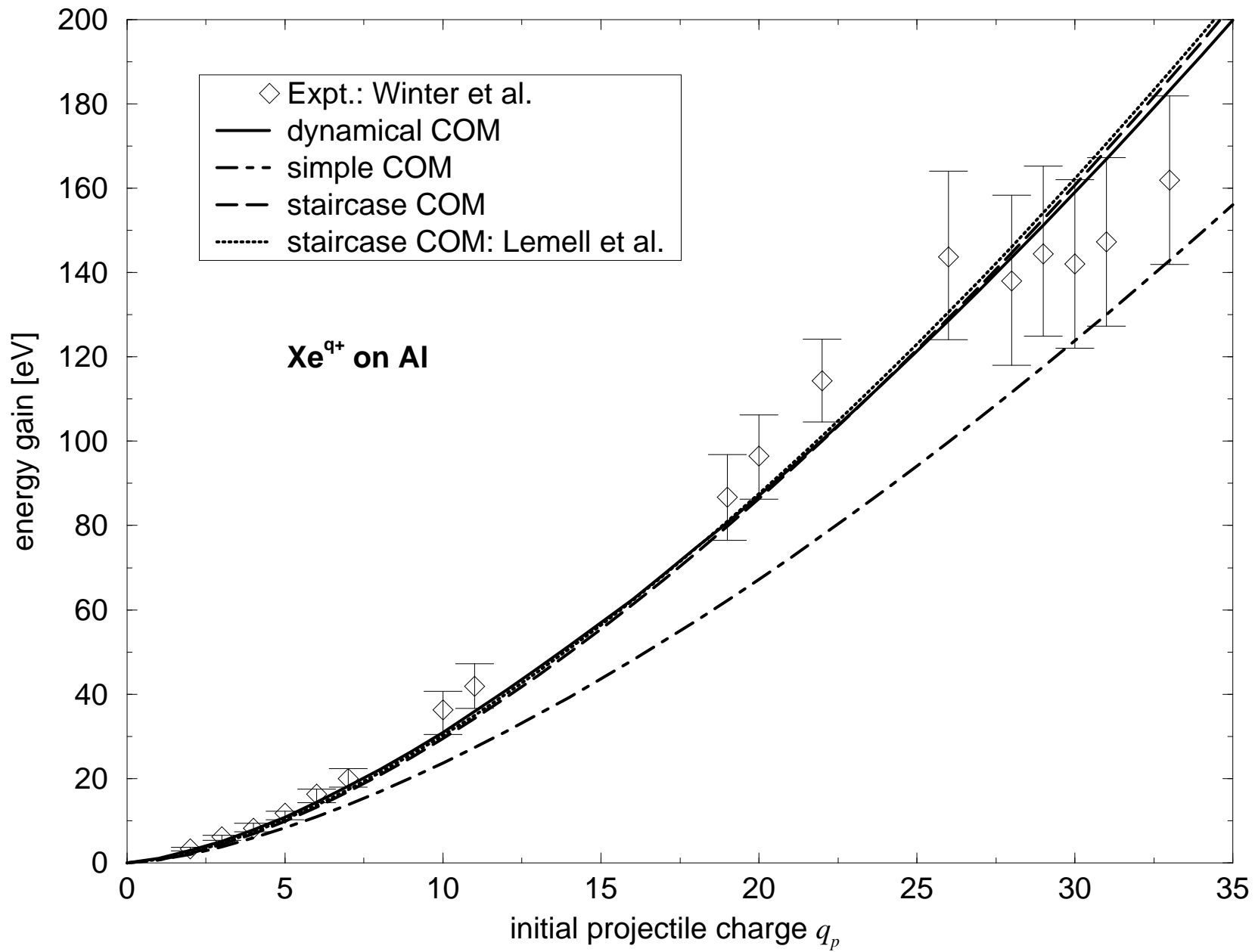
\begin{figure}
  \caption{Experimental \protect\cite{Win93}, simulated staircase COM
    results \protect\cite{Lem96}, and our simulated energy gains
    (dynamical COM) for $\mbox{Xe}^{q+}$ (3.7\,q~keV, $1.5^\circ$) on
    an Al surface. The simple model assumes instantaneous complete
    neutralization at the first critical distance $R_c$ and sets a
    lower boundary for projectile energy gains. The staircase COM
    instantaneously transfers one charge unit each time the
    over--barrier condition is fulfilled.  In the dynamical COM
    continuous charge currents flow between projectile and surface
    with rates derived from a classical model.}
  \label{fig:XeqqAl:Egain:sim:exp}
\end{figure}

The simple COM can be improved by letting one electron transfer to the
HCI each time the over--barrier condition is fulfilled at consecutive
critical radii for the first, second, etc.\ capture. This version of
the COM is called the {\it "staircase model"} \cite{Lem96}. In
contrast to the dynamical COM (Sec.~\ref{sec:COM}), in the simple and
staircase model the charge transfer current is quantized.  For energy
gains of Xe$^q+$ projectiles on an Al surface, the staircase model
almost coincides for all initial charges $q_p$ with the dynamical COM,
and the simple model predicts, as expected, lower energy gains for all
$q_p$ (Fig.~\ref{fig:XeqqAl:Egain:sim:exp}). Except for the highest
charge states, both, the staircase and dynamical COMs agree with the
experimental gains of Winter et al.\cite{Win93}, even though the more
elaborate dynamical COM employs transition rates that depend on the
width and depth of the potential saddle.  The simple model
underestimates the measured energy gain, except for the highest charge
states, where agreement with the experiment may be fortuitous. Except
for the simple COM, all simulations intersect the experimental error
bars for charge states $q_p \leq 30$. All simulations show the general
$q_p^{3/2}$--trend.

The deviation in the experimental data from the approximate
$q_p^{3/2}$--proportionality of the energy gain in all COM versions
above $q_p \simeq 26$ (Fig.~\ref{fig:XeqqAl:Egain:sim:exp}) has been
scrutinized by Lemell et al. \cite{Lem96}. The authors rule out both
saturation effects in the surface charge density fluctuations induced
by the HCI at $R=R_c$ and effects due to the parallel velocity of the
HCI. They conclude that the measured deviation from the
$q_p^{3/2}$--proportionality is due to incomplete screening of outer
shells at decreasing $R$ which, for high initial projectile charges,
leads to a faster decrease of the effective projectile charge and,
therefore, to a diminished increase in the energy gain as a function
of $q_p$.  However, as far as we know, the initial charge state $q_p$
at which the experimentally observed plateau appears has not yet been
reliably calculated within any COM (see also the review article of
Winter \cite{Win96} and references therein).
Figure~\ref{fig:XeqqAl:Egain:sim:exp} also shows that the   
staircase COM calculation of Lemell et al. \cite{Lem96} agrees with
our results.

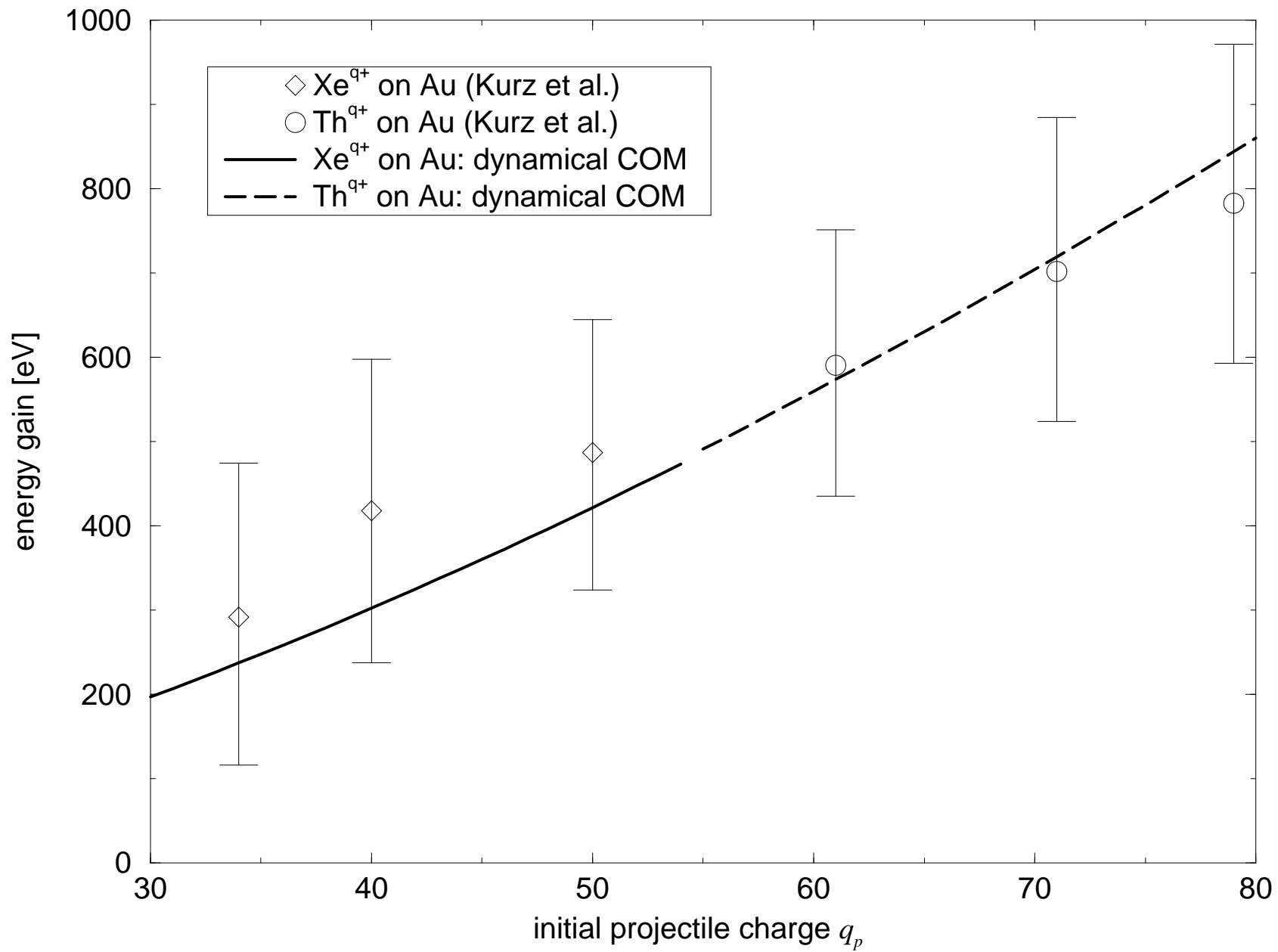
\begin{figure}
  \caption{Experimental \protect\cite{Kur94} and our 
    simulated data (dynamical COM) for very high charge state ions
    impinging on polycrystalline gold.}
  \label{fig:Xe:ThqqAu:Egain:sim:exp}
\end{figure}

Kurz et al. \cite{Kur94} have analyzed total electron yields for
higher charge states as a function of the inverse projectile velocity.
Their data for $\mbox{Xe}^{q+}, q=34\ldots 50$ and $\mbox{Th}^{q+},
q=61\ldots 79$ on gold surfaces under perpendicular incidence provide,
if at all, weak evidence for a deviation of the energy gain from the
$q^{3/2}$--proportionality (Fig.~\ref{fig:Xe:ThqqAu:Egain:sim:exp}).
We note that the experimental method of reference \cite{Kur94} is
prone to larger errors than the deflection angle method \cite{Win93}.
Our dynamical COM data for Thorium are near the upper end of the
experimental error bars. 

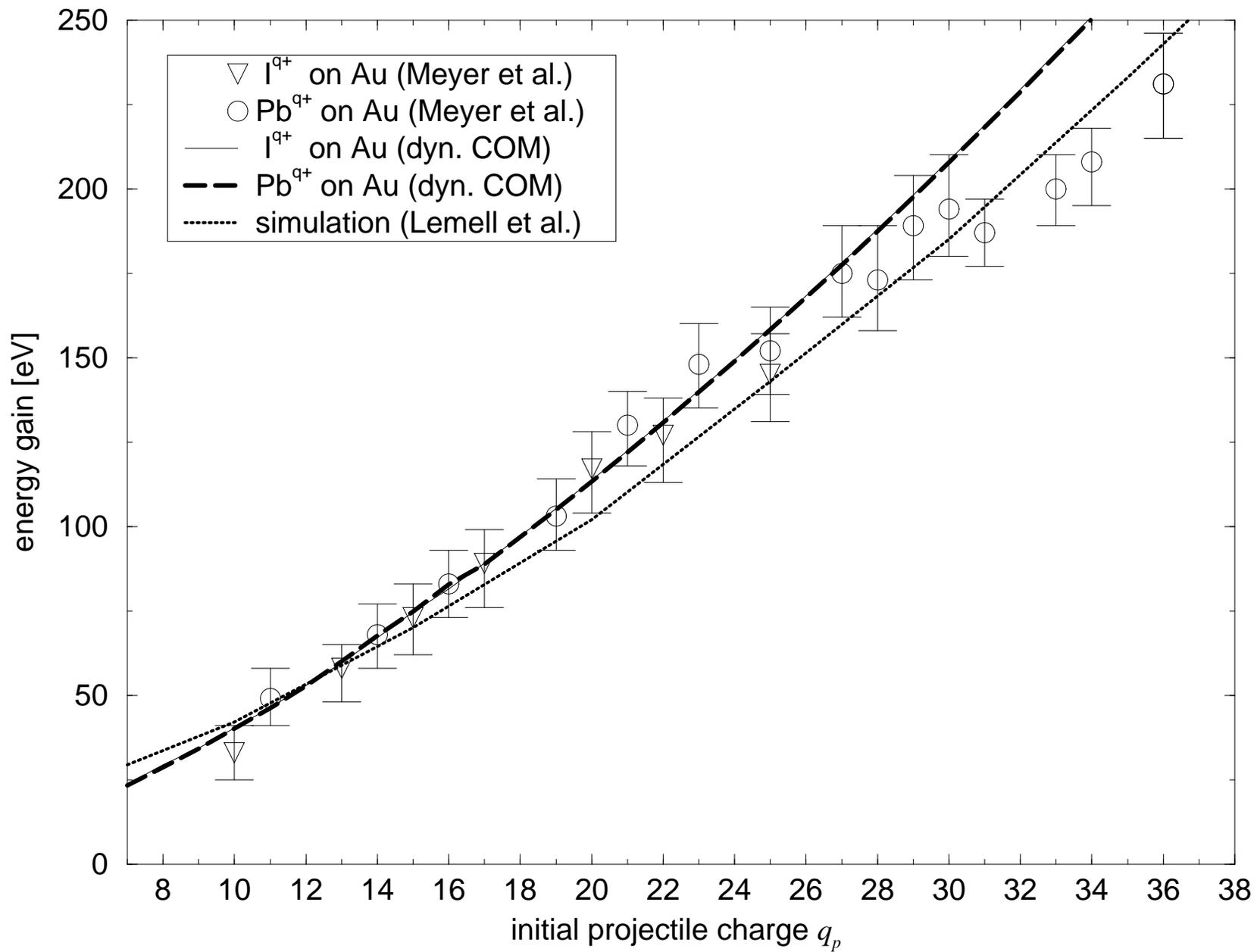
\begin{figure}
  \caption{Experimental \protect\cite{Mey95}, simulated staircase COM
    results \protect\cite{Lem96}, and our simulated data (dynamical
    COM) for 150~keV $\mbox{I}^{q+}$ and 
    $\mbox{Pb}^{q+}$ ions with charge states $q_p \leq 36$ on Au.} 
  \label{fig:I:PbqqAu:Egain:sim:exp}
\end{figure}

In Figure~\ref{fig:I:PbqqAu:Egain:sim:exp} we compare our dynamical
COM results for 150~keV ions on Au 
with energy gains measured by Meyer et al.\  \cite{Mey95}
and with the COM simulation of Lemell et al. \cite{Lem96}.  The
experimental results show good overall agreement with our calculations
for both projectiles but fall systematically short of the dynamical
COM values above $q_p \simeq 30$.

\subsubsection{Insulators}
\label{sec:insulators}

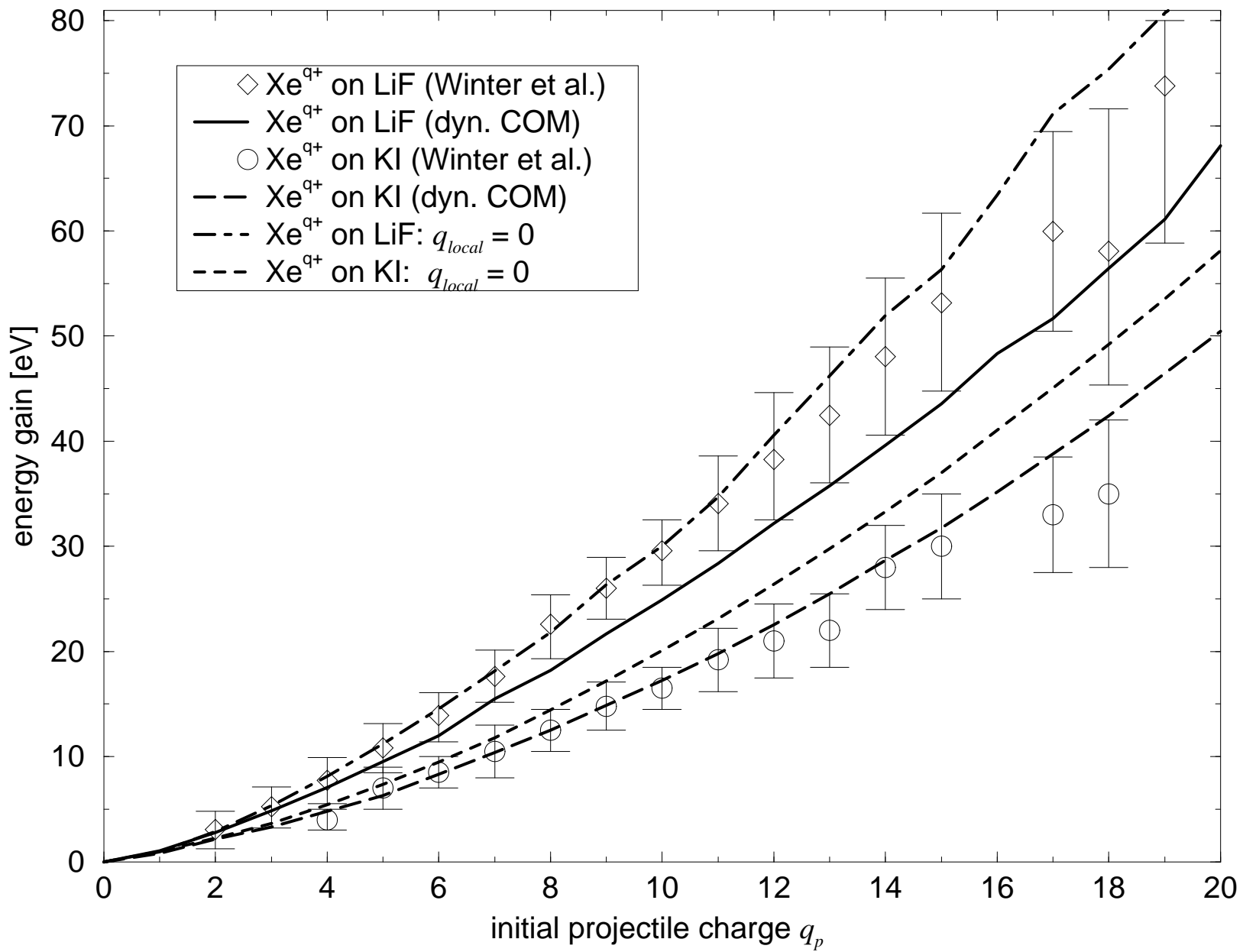
\begin{figure}
  \caption{Experimental energy gains \protect\cite{Aut96} compared with
    our dynamical COM simulations. Results obtained by neglecting
    capture--induced local surface charges are labeled as $q_{local}=0$.}
  \label{fig:XeqqLiF:KI:Egain:sim:exp}
\end{figure}

Energy gains for 50~keV Xe ions directed under a grazing incidence
angle of 1$^\circ$ on alkali halide crystals (LiF and KI) have been
measured recently by Auth et al. \cite{Aut95,Aut96} by using the
deflection angle method \cite{Win93}. Our extended dynamical COM
simulations agree with experiment for the KI target for $q_p<17$
(Fig.~\ref{fig:XeqqLiF:KI:Egain:sim:exp}).  However, experiment and
simulation tend to deviate in a systematic way for the LiF target,
where, for low and intermediate incident charge states, the measured
values slightly exceed our simulations.  We tried to identify an
adjustable parameter in order to further improve the agreement with
experiment for both targets.  In the oncoming paragraphs we will
discuss several of the effects that appear in our insulator extension
of the COM and use $\mbox{Xe}^{15+}$ on LiF and KI, with an energy
gains of 43.6~eV and 31.8~eV, as a reference.

At first we take a closer look at effects that are induced by the
surface charges. We observe that the restriction of one removable
charge per $\mbox{F}^-$ ion (which we did not apply to iodine for its
vast number of outer shell electrons) lowers the energy gain by
1.6~eV. Furthermore, disregarding all capture--induced surface charges
(cf.\ curves labeled $q_{local}=0$ in
Fig.~\ref{fig:XeqqLiF:KI:Egain:sim:exp}) increases the energy gain of
$\mbox{Xe}^{15+}$ on LiF to 56.3~eV.  For the KI target, however, the
neglect of surface charges increases the energy gain to 37.0~eV for
incident $\mbox{Xe}^{15+}$ which lies above the experimental error
bars.  Our numerical results show the expected increase for energy
gains at all charge states if we discard surface charges.  For the KI
target, the inclusion of capture--induced surface charges improves the
agreement between simulated and measured energy gains.

With respect to the ionic conductivities, we note that $\sigma$ has to
be increased by more than six orders of magnitude in order to induce
any significant change in the energy shifts. Despite our crude
estimates for $\sigma$ and the order--of--magnitude derivation of the
time constant $\tau$ in (\ref{equ:qtau}), we can therefore exclude
life--time effects of capture--induced surface charges on the
simulated energy gains.

The image plane is located at one anionic radius above the uppermost
bulk layer. This choice constitutes an upper limit for $z_{im}$.
Alternatively, as an accurate value is difficult to assess and the
concept of an image plane is not well defined for ionic crystals, one
could use $z_{im}$ as an adjustable parameter.  Placing the charge
distribution at the topmost lattice plane diminishes the image energy
for $\mbox{Xe}^{15+}$ by 1.6~eV on LiF.  For KI the energy gain
slightly increases by 0.6~eV.

We have performed all simulations with free--electron densities of
states and a constant volume factor ${\it V}=1$ in the transition
rates (\ref{equ:I}), which is a rather poor approximation for an ionic
crystal and may not give sufficient credit to the characteristics of a
particular crystal; the negative fluorine ion possesses six
$2p$--electrons, whereas iodine ion holds a large number of loosely
bound electrons.  In an attempt to work this information into the
simulation, we reduced the resonant gain rates (\ref{equ:I}) for the
LiF target by a factor of 5.  As a result we find that these modified
rates lead to energy gains on LiF that lie inside the experimental
error bars for all charge states.  In a more realistic representation
of the target electronic structure, more attention must be given to
the valence electrons of the anions.  The above--mentioned
discrepancies for KI and higher charges of the incident projectile may
be related to the simplified representation of the target electronic
structure inherent in our implementation of the dynamic COM.
 
As explained in Section IIA, our simulations were limited to
trajectories with $Y_p=0$, for which the collision plane intersects
anionic and cathionic nuclei along the [100] direction. Since in
surface scattering experiments the incident ion beam 
illuminates a surface area that is large compared with a surface unit
cell, we addressed the sensitivity of our simulated energy gains to
changes in $Y_p$. For $Y_p=0.5 d$, corresponding to surface--projected
trajectories half way between ionic rows, we find for 30~keV
$\mbox{Xe}^{15+}$ projectiles incident under 1$^\circ$ on LiF a
kinetic energy gain of 44.6~eV, compared to 43.6~eV for trajectories
with $Y_p=0$ and an experimental value \cite{Aut96} of 53.2~eV. The
slightly larger energy gain is consistent with the increased average
distance of capture--induced surface charges from the projectile. The
change in energy gain with $Y_p$ is sufficiently small such that,
within the overall accuracy of our calculation, we do not need to
include a time--consuming average over trajectories with different
$Y_p$ in our simulation.

\section{Conclusions}
\label{sec:conclusions}

In this work we applied and discussed extensions to the classical
over--barrier model that include insulator specific effects such as
capture--induced local surface charges, local workfunction changes,
and the dielectric response of the target. A detailed study of the
interaction mechanism has been presented in terms of the time
evolution of projectile level occupations, transition rates, and
several other quantities involved in the neutralization process.

Our results are in good agreement with previously published
experimental data for highly charged ions impinging on two different
alkali halide ionic crystals. In order to verify the basic framework
of our implementation of the dynamical COM, we have disabled all
effects related to insulators and found good agreement with energy
gain measurements for a variety of incident ion charge states and
metal targets.

\vspace{.20in}
\noindent {\large\bf Acknowledgments} \newline

We acknowledge helpful discussions with  H.J. Andr\"a, L. H\"agg, and
C.O. Reinhold.
This work was supported by the Division of Chemical Sciences, Basic
Energy Sciences, Office of Energy Research, U.S. Department of Energy,
by the Kansas Center for Advanced Scientific Computing sponsored
by the NSF EPSCoR/K*STAR program, and by the National Science
Foundation. One of us (JJD) is supported in part by the German
Bundesministerium f\"ur Bildung, Wissenschaft, Forschung und
Technologie under contract no. 13N6776/4.

\vspace{2cm}


\begin{tabular} {p{6.1in}l}
  Table I. Static limit ($\epsilon_0$), optical limit
  ($\epsilon_\infty$) and characteristic frequency $\omega_0$ used in
  (\ref{equ:epsilon}) for the dielectric response of LiF and KI
  crystals at T=$290^\circ$C \cite{Low69}.
  \label{tab:diel}
\end{tabular}

\vspace{5mm}

\begin{center}
\begin{tabular}{|c|c|c|c|} \hline
 & $\epsilon_0$ & $\epsilon_\infty$ & $\omega_0$ in 10$^{-3}$ a.u. \\ \hline
    LiF & 9.00 & 1.93 & 1.39 \\ \hline
    KI  & 5.09 & 2.65 & 0.46 \\ \hline
\end{tabular}
\end{center}  


\begin{references}
\bibitem{And91} H.J. Andr\"a {\it et~al.}, in {\em Proceedings of the
    XVII International Conference on the Physics of Electronic and
    Atomic Collisions}, Brisbane, Australia, 1991, edited by W.R.
  MacGillivray, I.E. McCarthy and M.C. Standages (Adam Hilger, N.Y.
  1992).
\bibitem{Bur93} J. Burgd\"orfer, in {\em Review of Fundamental
    Processes and Applications of Atoms and Ions}, edited by C.D. Lin
  (World Scientific, Singapore, 1993).
\bibitem{Aum95} F. Aumayr, in {\em Proceedings of the XIX.
    International Conference on the Physics of Electronic and Atomic
    Collisions}, Whistler, Canada, 1995, AIP Conference Proceedings
  306 (AIP Press, Woodbury, N.Y. 1995).
\bibitem{Bur91} J. Burgd\"orfer, P. Lerner, and F. Meyer, Phys. Rev. A
  {\bf 44}, 5674 (1991).
\bibitem{Bur95} J. Burgd\"orfer, C. Reinhold, and F. Meyer, Nucl.
  Instrum. Methods Phys. Res. Sect.  B {\bf 98}, 415 (1995).
\bibitem{Bar95} A. B\'ar\'any and C.J. Setterlind, Nucl. Instrum.
  Methods Phys. Res. Sect.  B {\bf 98}, 184 (1995).
\bibitem{Thu95A} U. Thumm, J. Phys. B {\bf 27}, 3515 (1994); {\bf 28},
  91 (1995).
\bibitem{Thu95B} U. Thumm, T. Ba\c{s}tu\v{g}, and B. Fricke, Phys.
  Rev. A {\bf 52}, 2955 (1995).
\bibitem{Bur96} J. Burgd\"orfer, C. Reinhold, L. H\"agg, and F. Meyer,
  Aust. J. Phys. {\bf 49}, 527 (1996).
\bibitem{Lem96} C. Lemell, H.P. Winter, F. Aumayr, J.Burgd\"orfer, and
  F. Meyer, Phys. Rev. A {\bf 53}, 880 (1996).
\bibitem{Thu97} U. Thumm, Phys. Rev. A {\bf 55}, 479 (1997).
\bibitem{Bur87} J. Burgd\"orfer, E. Kupfer, and H. Gabriel, Phys. Rev.
  A {\bf 35}, 4963 (1987).
\bibitem{Thu89}
U. Thumm, J. Phys. B {\bf 25},  421  (1992); U. Thumm and J.S. Briggs,
  Nucl. Instrum. Methods Phys. Res. Sect.  B {\bf 40/41},  161  (1989);
  {\bf 43}, 471 (1989); {\bf 47} 476 (1990).
\bibitem{Bor93} A. Borisov, D. Teillet-Billy, and J. Gauyacq, Nucl.
  Instrum. Methods Phys. Res. Sect.  B {\bf 78}, 49 (1993).
\bibitem{Deu97} S.A. Deutscher, X.Yang, and Burgd\"orfer, Phys. Rev. A
  {\bf 55}, 466 (1997).
\bibitem{Wil95} U. Wille, Nucl. Instrum. Methods Phys. Res. Sect. B
  {\bf 100}, 303 (1995).
\bibitem{Bor96} A. Borisov, R. Zimny, D. Teillet-Billy, and J.
  Gauyacq, Phys. Rev. A {\bf 53}, 2457 (1996).
\bibitem{Kur96} P. K\"urpick and U. Thumm, Phys. Rev. A {\bf 54}, 1487
  (1996).
\bibitem{Nor96} P. Nordlander, Phys. Rev. B {\bf 53}, 4125 (1996).
\bibitem{Aum93} F. Aumayr, H. Kurz, D. Schneider, M.A. Briere, J.W.
  McDonald, C.E. Cunningham, and H.P. Winter, Phys. Rev. Lett. {\bf
    71}, 1943 (1993).
\bibitem{And93} H.J. Andr\"a, A. Simionovici, T. Lamy, A. Brenac, G. A.
  Pesnelle, Europhys. Lett.  {\bf 23}, 361 (1993).
\bibitem{Mey93} F.W. Meyer, C.C. Havener, and P.A. Zeijlmans van
  Emmichoven, Phys. Rev. A {\bf 48}, 4476 (1993).
\bibitem{Lim94a} J. Limburg, J. Das, S. Schippers, R. Hoekstra, and R.
  Morgenstern, Phys. Rev. Lett. {\bf 73}, 786 (1994).
\bibitem{Gre95} M. Grether, A. Spieler, R. K\"ohrbr\"uck, and N.
  Stolterfoht, Phys. Rev. A {\bf 52}, 426 (1993).
\bibitem{Bri90} J.-P. Briand, L. de Billy, P. Charles, S. Essabaa, P.
  Briand, R. Geller, J.-P. Declaux, S. Bliman, and C. Ristori, Phys.
  Rev. Lett. {\bf 65}, 159 (1990).
\bibitem{Sch91} M. Schulz, C. Cocke, S. Hagmann, M. St\"ockli, and H.
  Schmidt--Boecking, Phys. Rev. A {\bf 44}, 1653 (1991).
\bibitem{Win96A} H. Winter, J. Phys: Condens. Matter {\bf 8}, 10149
  (1996).
\bibitem{Mey95} F.W. Meyer, L. Folkerts, H.O. Folkerts, and S.
  Schippers, Nucl. Instrum. Methods Phys. Res. Sect. B {\bf 98}, 443
  (1995).
\bibitem{Win96} S. Winecki, C. Cocke, D. Fry, and M. St\"ockli, Phys.
  Rev. A {\bf 53}, 4228 (1996).
\bibitem{Yan96} Q. Yan, D.M. Zehner, F.W. Meyer, and S. Schippers,
  Phys. Rev. A {\bf 54}, 641 (1996).
\bibitem{Aut95} Ch. Auth, T. Hecht, T. Igel, and H. Winter, Phys. Rev.
  Lett. {\bf 74}, 5244 (1995).
\bibitem{Lim95} J. Limburg, S. Schippers, R. Hoekstra, R. Morgenstern,
  H. Kurz, F. Aumayr, and HP. Winter, Phys. Rev. Lett. {\bf 75}, 217
  (1995).
\bibitem{Aut96} Ch. Auth and H. Winter, Phys. Lett. A {\bf 217}, 119
  (1996).
\bibitem{Lim96} J. Limburg, S. Schippers, R. Hoekstra, R.
  Morgenstern, H. Kurz, M. Vana, F. Aumayr, HP. Winter, Nucl.
  Instrum. Methods Phys. Res. Sect. B {\bf 115}, 237 (1996).
\bibitem{Bor96A} A.G. Borisov, V.Sidis, and H.Winter, Phys. Rev.
  Lett. {\bf 77}, 1893 (1996).
\bibitem{Hag97} L. H\"agg, C.O. Reinhold, and J. Burgd\"orfer,
 Phys. Rev. A {\bf 55}, 2097 (1997).
\bibitem{Cas96} F. Casali and U. Thumm, Bull. Am. Phys. Soc.
 {\bf 41}, 1129 (1996).
\bibitem{Bar85} A. B\'{a}r\'{a}ny, G. Astner, H. Cederquist, H.
  Danared, S. Huldt, P. Hvelplund, A. Johnson, H. Knudsen, L. Liljeby,
  and K.-G. Rensfeld, Nucl. Instrum. Methods Phys. Res. B {\bf 9}, 397
  (1985).
\bibitem{Nie86} A. Niehaus, J. Phys. B {\bf 19}, 2925 (1986).
\bibitem{Ech92} F.J. Garc\'{\i}a de Abajo and P.M. Echenique,
  Phys. Rev. B {\bf 46}, 2663 (1992).
\bibitem{Low69} R. Lowndes and D. Martin, Proc. Roy. Soc. London A,
  473 (1969).
\bibitem{Jac75} J. Jackson, {\it Classical Electrodynamics}, 2nd ed.
  (Wiley, N.Y. 1975).
\bibitem{Jen88} P.J. Jennings, R.O. Jones and W.Weinert, Phys. Rev. B
  {\bf 37}, 6113 (1988).
\bibitem{Mot38} N.F. Mott and M.J. Littleton, Trans. Faraday Soc. {\bf
  34}, 485 (1938).
\bibitem{Mah80} G.D. Mahan, Phys. Rev. B {\bf 21}, 4791 (1980).
\bibitem{Pia81} M. Piacentini and J. Anderegg, Solid State Commun.
      {\bf 38}, 191 (1981).
\bibitem{Cow81} R. Cowan, {\em The Theory of Atomic Structure and
    Spectra} (University of California Press, Berkeley, 1981).
\bibitem{Ash76} N.W. Ashcroft and N.D. Mermin, {\it Solid State Physics}
  (W.B. Saunders, Philadelphia 1976).
\bibitem{OBr74} W.P. O'Brien, Jr. and J.P. Hernandez, Phys. Rev. B {\bf 8},
  3560 (1974).
\bibitem{Mol47} G. Moli\`ere, Z. Naturforschg.{\bf 2a}  133 (1949).
\bibitem{Gem74} D.~S. Gemmell, Rev. Mod. Phys. {\bf 46}, 129ff (1974).
\bibitem{Win93} H. Winter, Ch. Auth, R. Schuch, and E. Beebe, Phys. Rev.
  Lett. {\bf 71}, 1939 (1993).
\bibitem{Bur93A} J. Burgd\"orfer and F. Meyer, Phys. Rev. A {\bf 47},
  R20 (1993).
\bibitem{Kur94} H. Kurz, F. Aumayr, HP. Winter, D. Schneider, M.A.
  Briere, and J.W. McDonald, Phys. Rev. A {\bf 49}, 4693 (1994).
\end{references}
\end{document}